\documentclass[10pt]{article}
\usepackage{latexsym,amsmath,amssymb,graphics,amscd}
\usepackage{multirow}
\usepackage{subfigure}
\usepackage{natbib}
\usepackage{booktabs}
\usepackage{tabularx}
\usepackage[hang,footnotesize]{caption}
\usepackage[pdftex]{graphicx}
\usepackage{graphicx}
\usepackage{color}
\usepackage{setspace}
\usepackage{color}

\usepackage[pdftex,breaklinks,colorlinks,
  citecolor=blue,
  urlcolor=blue]{hyperref}

\addtocounter{footnote}{1}
\makeatletter
\@addtoreset{figure}{section}
\def\thefigure{\@arabic\c@figure}
\def\fps@figure{h,t}
\@addtoreset{table}{bsection}
\def\thetable{\@arabic\c@table}
\def\fps@table{h, t}
\@addtoreset{equation}{section}

\makeatother

\newtheorem{theorem}{Theorem}[section]

\newtheorem{proposition}[theorem]{Proposition}

\newsavebox{\savepar}

\makeatletter

\begin{document}

\title{\textbf{Quadratic hedging schemes for non-Gaussian GARCH models}}
\author{Alexandru Badescu$^{1}$, Robert J. Elliott$^{2,3}$, and Juan-Pablo Ortega$^{4}$ }
\date{}
\maketitle

\begin{abstract}
We propose different schemes for option hedging when  asset returns are  modeled using a general class of GARCH models. More specifically, we implement local risk minimization and a minimum variance hedge approximation based on an extended Girsanov principle that generalizes Duan's \citeyearpar{duan:GARCH:pricing} delta hedge. Since the minimal martingale measure fails to produce a probability measure in this setting, we construct local risk minimization  hedging strategies with respect to a pricing kernel. These approaches are investigated in the context of non-Gaussian driven models. Furthermore, we analyze these methods for non-Gaussian GARCH diffusion limit processes and link them to the corresponding discrete time counterparts. A detailed numerical analysis based on S\&P 500 European Call options is provided to assess the empirical performance of the proposed schemes. We also test the sensitivity of the hedging strategies with respect to the risk neutral measure used by recomputing some of our results with an exponential affine pricing kernel.
 \end{abstract}

\medskip



\makeatletter
\addtocounter{footnote}{1} \footnotetext{%
Department of Mathematics and Statistics, University of Calgary, Calgary, Canada.}
\addtocounter{footnote}{1} \footnotetext{%
Corresponding author: School of Mathematical Sciences, University of Adelaide, SA 5005, Australia, Phone: +61 8 8313 5077, Fax:  +61 8 8313 3696. E-mail: {\texttt{relliott@ucalgary.ca} }}
\addtocounter{footnote}{1} \footnotetext{%
Department of Mathematics and Statistics, University of Calgary, Calgary, Canada.}
\addtocounter{footnote}{1} \footnotetext{%
Centre National de la Recherche Scientifique, D\'{e}partement de Math\'{e}%
matiques de Besan\c{c}on, Universit\'{e} de Franche-Comt\'{e}, UFR des
Sciences et Techniques. 16, route de Gray. F-25030 Besan\c{c}on cedex.
France.}
\makeatother

\newpage

\section{Introduction}
\label{Introduction}

Many empirical studies have shown strong evidence against some of the underlying assumptions of the Black-Scholes~\citeyearpar{BS} model, in particular the constant value that is assumed for the volatility. The incorporation of stochastic volatility in modeling the dynamics of asset returns plays a major role in explaining some of the stylized properties of financial time series. In particular, the pricing and hedging of options based on stochastic volatility (SV) models have been frequently studied in  both discrete and continuous-time.
The discrete-time literature has been dominated by the Generalized Autoregressive Conditionally Heteroskedastic (GARCH) models introduced by Engle~\citeyearpar{engle:arch} and Bollerslev~\citeyearpar{bollerslev:garch}. Although these models have been extensively investigated for pricing,
they have rarely been used for option hedging purposes.\footnote{A very recent survey of the major contributions to GARCH option pricing has been provided by Christoffersen \textit{et al.}~\citeyearpar{CJO13}.}

The first proposal in this direction can be found in Duan~\citeyearpar{duan:GARCH:pricing} who introduced a delta hedge obtained as the partial derivative of the option price with respect to the price of the underlying and computed it using Monte Carlo simulations in order to evaluate the corresponding locally risk neutralized valuation relationship (LRNVR) expectation.  Other approaches for computing this hedge ratio include the finite difference method  proposed by Engle and Rosenberg~\citeyearpar{ER95} or the adjusted Black Scholes delta hedging computations based on analytic approximations of GARCH option prices (see e.g.  Choi~\citeyearpar{Choi}). Kallsen and Taqqu~\citeyearpar{kallsen:taqqu} proposed a continuous time  GARCH process  under  which the associated markets are complete and also investigated hedging in the Black-Scholes sense. Duan's~\citeyearpar{duan:GARCH:pricing} definition of the delta hedge has been subjected to criticisms, the most important coming from  Garcia and Renault~\citeyearpar{garcia:renault} who pointed out that this specification ignores the fact that the option price depends also on the conditional variance, which is itself a function of the asset price. They showed that when the option price is homogeneous of degree one with respect to the asset price and the strike price, then the option delta hedge is consistent with Duan's formulation. Further  discussions on  homogeneity of option prices and delta hedging for scale-invariant models are provided in Bates~\citeyearpar{Bates} and Alexander and Nogueira~\citeyearpar{AN07a, AN07b}. In order to take into account the stochastic volatility environment of GARCH models, Engle and Rosenberg~\citeyearpar{ER00} considered the computation of the option's deltas and gammas using the chain rule and studied the modifications in empirical performance when hedging options with changing variance term structure for different volatility specifications.

The techniques briefly described above try to extend the hedging  designed for complete markets  to a GARCH context, and therefore are not  optimal in the incomplete market setting, that is, the use of continuous time hedging ratios for discrete-time models may generate additional hedging errors. Thus, various other  hedging techniques have been proposed in the financial literature by minimizing different measures for the hedging error. One of the most influential strategies in this direction has been the local risk minimization scheme introduced by F\"ollmer, Schweizer, and Sondermann (see~\citeyearpar{foellmer:sondermann, foellmer:schweizer, schweizer:2001}, and references therein); this approach uses the notion of sequential regressions in order to construct a generalized trading strategy (not necessarily self-financing) that minimizes daily squared hedging errors. Local risk minimization is particularly convenient in the discrete time setup because it is well adapted to recursive  algorithmic implementations and can be tuned to minimize hedging errors related to arbitrary prescribed hedging frequencies. The standard approach for constructing local risk minimization hedges is to  minimize the conditional  expected value of the quadratic cost of hedging under the historical probability measure. This optimization procedure also gives  rise to a contingent claim price computed in the risk neutral world with the well-known minimal martingale measure. Although local risk minimization has been widely used for pricing and hedging financial derivatives based on continuous time models, its implementation in discrete-time, and in particular for GARCH models, raises some issues. For example, in Ortega~\citeyearpar{ortega:garch:pricing} it has been shown that in the GARCH context the minimal martingale measure is signed, and it is only under very restrictive hypotheses (the model innovations are required to be bounded) that option prices arising from local risk minimization using the physical measure are guaranteed to be arbitrage free.  Additionally, a number of situations have been identified (see e.g. {\v{C}}ern{\'y} and Kallsen~\citeyearpar{cerny:kallsen:annals}) in which the local minimization of squared hedging errors does not guarantee the minimization of the final unconditional expected squared hedging error; this point has motivated recent important work which targets the minimization of the global hedging risk (see  e.g. {\v{C}}ern{\'y} and Kallsen~\citeyearpar{cerny:kallsen:annals, cerny:kallsen:counterexample, cerny:kallsen}). Therefore, one has to consider the local quadratic minimization in a different context.

In this paper we provide a unified framework for the local risk minimization and the standard minimum variance delta hedges for a general class of asymmetric non-Gaussian GARCH models. Firstly, we introduce local risk minimization with respect to a martingale measure for different hedging frequencies under a GARCH model with an unspecified distribution for the driving noise. When local risk minimization is carried out with respect to a martingale measure, a number of interesting features arise. For example, the option prices that result from the scheme are guaranteed to be arbitrage free and the hedging strategies in this case minimize, not only the local but also the global hedging risk. Additionally, there is a decoupling between the expressions that determine hedges and prices that can be taken advantage of in order to create numerically performing schemes to compute them.  An intuitive explanation behind risk neutral optimization and references to other research studies which undertake this approach are presented in Cont \textit{et al.}~\citeyearpar{Cont}. To our knowledge, the only attempts at exploring  local risk minimization in a GARCH setting are due to Ortega~\citeyearpar{ortega:garch:pricing} and Huang and Guo~\citeyearpar{huang:guo:2}. The former investigates the use of the local risk minimization with respect to a martingale measure in the context of Gaussian driven GARCH processes and shows how Duan~\citeyearpar{duan:GARCH:pricing} and Heston and Nandi's~\citeyearpar{heston:nandi} pricing formulas can be recovered in this fashion, while the latter investigates the risk adjusted expected quadratic cost of hedging minimal hedging strategies in the sense of Elliott and Madan~\citeyearpar{elliott:madan:mcmm}. 


When the time between two consecutive observations approaches zero, we show that the local risk minimization hedge ratio under a martingale measure can be approximated with a stochastic volatility type delta hedge. More specifically, the resulting delta consists of the sum of the option delta (the first partial derivative of the option price with respect to the spot price) and the option vega (the first partial derivative of the option price with respect to the conditional  variance process) scaled by a so called vega multiplier.  The name of this term is borrowed from Engle and Rosenberg~\citeyearpar{ER00}. However, this approximation is different to the one proposed in that paper since all our Greeks and vega multipliers are computed directly based on a GARCH model instead of on an adjusted Black-Scholes price in the sense of Hull and White~\citeyearpar{HULL1987}. Using the extended Girsanov principle of Elliott and Madan~\citeyearpar{elliott:madan:mcmm}  we derive simplified expressions for these quantities under the general non-Gaussian setup, although the explicit computation of Greeks involves Monte Carlo simulation. This result generalizes the static delta hedge proposed by Duan~\citeyearpar{duan:GARCH:pricing}. In fact, the latter appears as a particular case of the one that we propose if changes in asset prices and conditional variances are uncorrelated.  Finally we derive the local risk minimizing hedges under the non-GARCH diffusion limits of our model and show their relationships with our proposed discrete-time counterparts. This result also serves as a further justification of why Duan's~\citeyearpar{duan:GARCH:pricing}  hedge is the correct instantaneous delta  hedge only under the aforementioned assumption. 

We provide an extensive numerical example in which we test the out-of-sample empirical performance of the proposed delta hedges for a special case of GARCH model driven by Normal Inverse Gaussian distributed innovations in hedging a collection of European call options on the  S\&P 500 index. The GARCH model parameters are estimated via maximum likelihood based on the historical returns on the underlying S\&P 500 index. The benchmark hedge ratio used in our analysis  is the Ad-hoc Black Scholes model proposed by  Dumas \textit{et al.}~\citeyearpar{DumasAdhoc}. Our results show that overall both the local risk minimization strategy and its delta approximation outperform the Ad-hoc Black-Scholes benchmark. A detailed analysis of the hedging errors is provided for different groups of moneyness and times to maturity. Finally, we test the sensitivity of the schemes that we propose with respect to the choice of the pricing measure by using an exponential affine stochastic discount factor. As a result, we see that this other martingale measure leads to similar hedging performances and that there are no significant differences between the results obtained with the two pricing kernels.

The rest of the paper is organized as follows. Section~\ref{Section2} discusses  local risk minimization with respect to the physical and risk neutral measures, and their relationship with minimum variance hedging.  In Section~\ref{LRMandDelta} we compute quadratic hedges for general discrete time GARCH models and their diffusion limits. Section~\ref{NIGSection} investigates the empirical performance of the proposed GARCH hedges using both asset returns and option data.  Section~\ref{Conc} concludes the paper.

\section{Local risk minimization for GARCH options}
\label{Section2}

Consider a discrete time financial market with the set of trading dates  indexed by $\mathcal{T} = \{t | t = 0,\dots,T\}$. The market consists of one reference and one risky asset. We denote by $P$ the underlying probability measure and we assume that the dynamics of these assets are modeled by the following bivariate stochastic process $\left(S^0, S\right) =  \left\{\left(S^0_t, S_t\right)\right\}_{t \in \mathcal{T}}$.  
The risk-free asset is assumed to follow a deterministic process and it evolves over time according to the following equation:
\begin{eqnarray}
S^0_t = e^{rt},\;\;\;\;S^0_0 = 1 .\label{bond}
\end{eqnarray}
Here $r$ represents the instantaneuous  risk free rate of return.
The log-return on the risky asset process, denoted by $y = \left\{y_t\right\}_{t \in \mathcal{T}} : =  \left\{\ln \left(S_t / S_{t-1}\right)\right\}_{t \in \mathcal{T}}$, has a general asymmetric GARCH(1,1) structure (NGARCH) with the $P$-dynamics given below:
\begin{eqnarray}
y_t   & = &  \mu_t + \sigma_t \epsilon_t,\;\;\;\;\epsilon_t |\mathcal{F}_{t-1} \sim \mathbf{D}(0,1), \label{retPg} \\
\sigma^2_t & = &  \alpha_0 + \alpha_1\sigma^2_{t-1}\omega\left(\epsilon_{t-1} \right) + \beta_1\sigma^2_{t-1} .  \label{volPg}
\end{eqnarray}
 The uncertainty modeled through the complete probability space $\left(\Omega, \mathcal{F}, P\right)$ has a filtration $\mathcal{F} := \{\mathcal{F}_t \}_{t \in {\cal T}}$, where $\mathcal{F}_t $ represents the $\mathcal{\sigma}$-field of all market information available up to time $t$. The innovations $\left\{\epsilon_t\right\}_{t \in \mathcal{T}} $ is a sequence of random variables such that for each $t$, $\epsilon _t $ is independent and identically distributed conditional on $\mathcal{F}_t $ with a standardized   distribution $\mathbf{D}(0,1)$ under $P$. 
 We assume that the innovations conditional cumulant generating function  exists in a neighborhood of zero, that is:
\[ \kappa_{\epsilon_t}(z) = \ln {E^P_{t-1} \left[ e^{z\epsilon_t} \right] }  < \infty, \;\;\;\;z \in (-u,u),\;\; u > 0 . \] 
The conditional mean log-return $\mu_t$ is $\mathcal{F}_{t-1}$  measurable. A specific form in the spirit of Christoffersen \textit{et al.}~\citeyearpar{Chris09b} will be used starting in the following section. The conditional variance equation has a GARCH(1,1) structure with a news impact curve modeled through the function $\omega(\cdot)$. Our theoretical and numerical results in the next sections will be derived based on an asymmetric NGARCH structure. Finally, the parameters are assumed to satisfy the standard second order stationarity conditions.
It is well known that, in the absence of arbitrage, the price of any contingent claim can be expressed as the discounted expected value of its payoff at maturity under a $P$-equivalent risk neutral probability measure. Since the market implied by GARCH models is incomplete, there exists an infinite number of such pricing measures.  In order to conclude the characterization of our GARCH(1,1) pricing model we need to specify the risk neutralized dynamics of the log-returns under such a measure. Thus, we propose below the following general GARCH(1,1)-type structure for the log-return process under a  martingale measure generically denoted by $Q$: 
\begin{eqnarray}
y_t  & = & r - \kappa^*_{\epsilon^*_t}(\sigma^*_t) + \sigma^*_t\epsilon^*_t,\;\;\;\; \epsilon^*_t |\mathcal{F}_{t-1} \sim \mathbf{D^*}(0,1) \label{retQ} \\
\sigma^{*2}_t & = & \alpha^*_{0t} +\alpha^*_{1t}\sigma^{*2}_{t-1}\omega^*(\epsilon^*_{t-1}) + \beta^*_{1t}\sigma^{*2}_{t-1} .  \label{volQ}
\end{eqnarray}
The above dynamics provide us with a  general GARCH characterization in the risk neutral world. This structure nests some of the most popular martingale measures used in the GARCH option pricing literature. In particular, the extended Girsanov principle and the exponential affine discount factors can be viewed as special cases of (\ref{retQ})-(\ref{volQ}). All risk neutral quantities are illustrated through a ``*'' notation, so that $\sigma^{*2}_t$ represents the risk neutral conditional variance of the log-return process, while $\kappa^*_{\epsilon^*_t}(\cdot)$ is the $\mathcal{F}_{t-1} $- conditional cumulant generating function of $\epsilon^*_t$ under $Q$. The standard innovation process under the martingale measure is assumed to have a conditional distribution $\mathbf{D^*}(0,1) $ under the martingale measure that is in general  different from $\mathbf{D}(0,1) $. However, in many practical examples $\mathbf{D^*}$ and $\mathbf{D}$ coincide.  Notice that the GARCH-type parameters may no longer be constants under the new measure. However, in order to keep the predictability structure of the conditional variance under $Q$, we assume that $\alpha^*_{0t}$, $\alpha^*_{1t}$, and $\beta^*_{1t}$ are all $\mathcal{F}_{t-1}$ measurable.  


\subsection{Local risk minimization with respect to the physical measure}
\label{Local risk minimization for ARSV options}

Since the market associated with GARCH modeled assets is incomplete, there exist contingent products  that cannot be fully replicated  using a self-financing portfolio made out of its underlying  and a bond. A number of techniques have been developed over the years that tackle simultaneously the pricing and hedging problems by replacing the concept of  replication with that of hedging efficiency. As we shall see, these techniques provide both expressions for prices and hedging ratios which, even though they require in most cases Monte Carlo computations, admit convenient interpretations based on the notion of hedging error minimization that can be adapted to various models for the underlying asset. In the following paragraphs we briefly describe the $P$-local risk minimization  strategy for hedging a European contingent claim $H$ that depends on the terminal value of the risky asset  $S _t $.

 A   generalized trading strategy is a pair of stochastic processes $(\xi^0, \xi)$ such that  $\{\xi^0_t\}_{t \in \{0, \ldots,T\}}$    is adapted and $\{\xi_t\} _{t \in \{1, \ldots,T\}}$ is predictable. The associated value process $V $ of $(\xi^0, \xi) $ is defined as
\begin{equation*}
V _0:= \xi_0^0, \quad \mbox{ and  } \quad V _t:= \xi_t^0 \cdot S ^0_t+ \xi_t\cdot S _t, \quad t=0, \ldots, T.
\end{equation*}
The  cost process $C$ corresponding to this strategy is defined by:
\begin{equation*}
C _t:=V _t - \sum_{k=1}^t \xi_k\cdot \left(S _k-S_{k-1}\right) , \quad t=0, \ldots, T.
\end{equation*}
The locally risk minimization hedging strategy  optimizes the following risk process:
\begin{equation*}
LR _t(\xi^0, \xi):=E^P \left[\left(\widetilde{C} _{t+1}-\widetilde{C} _t\right)^2\mid \mathcal{F}_t\right], \quad t=0, \ldots, T-1. 
\end{equation*}
Here $\widetilde{C}_t$ is the discounted cost price at time $0$, $\widetilde{C}_t = C_t e^{-rt}$. 
It has been  showed that,  whenever a local risk-minimizing technique exists, it is fully determined by the backwards recursions:
\begin{eqnarray}
V _T &= & H,\label{hedge general 1}\\
\xi^{P}_{t+1} &= & \frac{{\rm Cov}^P\left(\widetilde{V}_{t+1},\widetilde{S}_{t+1}-\widetilde{S }_t\mid \mathcal{F} _t\right)}{{\rm Var}^P\left[\widetilde{S}_{t+1}-\widetilde{S }_t\mid \mathcal{F} _t\right]}, \label{hedge general 2}\\
\widetilde{V} _t &= &E^P\left[\widetilde{V}_{t+1}\mid \mathcal{F} _t\right]- \xi_{t+1}E^P\left[\widetilde{S}_{t+1}-\widetilde{S }_t\mid \mathcal{F} _t\right].\label{hedge general 3}
\end{eqnarray}
The upper script in the hedging ratio formula (\ref{hedge general 2}) indicates that the solution is obtained by minimizing the local risk under $P$. 
An appealing feature of local risk minimization consists of its adaptivity to prescribed changes in the hedging frequency. Indeed, without the loss of generality we suppose that the life of the option $H$ with maturity in $T$ time steps is partitioned into identical time intervals of duration $j$; this assumption implies the existence of an integer $k$ such that  $kj=T $. We now wish to set up a  local risk minimizing replication strategy for $H$ in which hedging is carried out once every $j$ time steps. We shall denote by $\xi_{t+j} $ the hedging ratio at time $t$ that presupposes that the next hedging will take place at time $t+j $. The value of such ratios will be obtained by minimizing the $j$-spaced local risk process:
\begin{equation}
\label{local risk changed frequency}
LR _t^{(j)}(\xi^0, \xi):=E^P \left[(\widetilde{C} _{t+j}^{(j)}-\widetilde{C} _t^{(j)})^2\mid \mathcal{F}_t\right], \quad t=0, j, 2j,\ldots, (k-1)j=T-j .
\end{equation}
Here $\{\widetilde{C} _t^{(j)} \}$ is the cost process associated with this type of strategies. 
A convoluted but straightforward induction argument shows that the $j$-step local risk minimization is the solution of the following backward recursions:
\begin{eqnarray}
V^{(j)}_T &= & H, \label{hedge general frequency 1}\\
\xi^{P}_{t+j} &= & \frac{{\rm Cov}^P(e^{-rT}H\cdot N _T\cdot N_{T-j}\cdots N_{t+2j},\widetilde{S}_{t+j}-\widetilde{S }_t\mid \mathcal{F} _t)}{{\rm Var}^P[\widetilde{S}_{t+j}-\widetilde{S }_t \mid \mathcal{F}_t]},\label{hedge general frequency 2}\\
\widetilde{V}^{(j)} _t &= &E_t^P\left[e^{-rT}H\cdot N _T\cdot N_{T-j}\cdots N_{t+j}\right],\quad t=0,j, 2j,\ldots,T-j,\label{hedge general frequency 3}
\end{eqnarray}
Here, for any $t=j, 2j, \ldots, T$, the process $N_t$ is given by:
\begin{eqnarray}
\label{Radon-Nikodym derivative mmm}
N _t = \frac{1- \gamma_t(\widetilde{S}_t - \widetilde{S}_{t-j})} { E^P\left[ 1- \gamma_t\left(\widetilde{S}_t - \widetilde{S}_{t-j}\right) \mid \mathcal{F}_{t-j} \right]}  =1- \frac{E^P\left[\widetilde{S}_{t}-\widetilde{S }_{t-j} \mid \mathcal{F}_{t-j} \right]  \left(\widetilde{S}_{t}-E^P\left[\widetilde{S }_{t} \mid \mathcal{F}_{t-j} \right]\right)}{{\rm Var}^P\left[\widetilde{S}_{t}-\widetilde{S }_{t-j} \mid \mathcal{F}_{t-j}\right ]},
\end{eqnarray}
where, \[\gamma_t  = \frac {E^P \left[\widetilde{S}_{t}-\widetilde{S }_{t-j} \mid \mathcal{F}_{t-j} \right]}{ E^P\left[\left( \widetilde{S}_{t}-\widetilde{S }_{t-j}\right)^2 \mid \mathcal{F}_{t-j} \right] } .\]
When the product  $Z _T:=\prod_{t \in \{j, 2j,\ldots, T\}} N _t$ is a positive random variable, the relation $Z _T:= dQ^{min}/ dP$ defines a measure equivalent to $P$, called the  minimal martingale measure, for which the discounted partial process $\{\widetilde{S}_j, \widetilde{S}_{2j},\ldots, \widetilde{S}_T\}$ is a martingale. The interest of this measure is that, when it exists and in the case $j=1 $, the value process of the local risk-minimizing strategy  with respect to the physical measure coincides with arbitrage free price for $H$ obtained by using $Q^{min}$ as a pricing kernel. 


There are various implementation issues for the recursions~(\ref{hedge general frequency 1})--(\ref{hedge general frequency 3})  that need to be taken care of when asset returns are modeled by GARCH processes; we analyze in detail some of them in the next section where we also propose an alternative hedging scheme.

\medskip

\subsection{ Local risk minimization with respect to a martingale measure} \label{LRMQ}

The minimal martingale measure is usually a signed measure in discrete time settings, so derivative valuation based on this approach might not be suitable. For example, the Radon-Nikodym process defined in~(\ref{Radon-Nikodym derivative mmm}) can take negative values and this may lead to negative prices when computing the expectations from~(\ref{hedge general frequency 3}) using Monte Carlo simulation. For asset returns governed by GARCH models,  one can impose conditions on the parameter set or the innovations which can be too restrictive and make the estimation more challenging. 
A sufficient condition in the GARCH context requires the driving noise to take bounded values (see, for example,~\citet{ortega:garch:pricing}).\footnote{An unreported numerical simulation study indicates that negative values for $N_t$ happen very rarely for GARCH processes driven by Gaussian innovations. This is no longer the case for a variety of other heavier tailed densities whose exploration is one of the main objectives of the current paper.} 

The recursive derivation of a derivative price according to ~(\ref{hedge general frequency 1})-(\ref{hedge general frequency 3}) is not an easy task even in settings when the minimal martingale measure is a probability measure.  Both the option prices  and hedging ratios computations require the use of Monte Carlo simulations under the physical measure $P$.
An inspection of the expressions~(\ref{hedge general frequency 1})--(\ref{hedge general frequency 3}) shows that in order to compute them we need first an estimation of the stochastic discount factors $\{N _j, \ldots, N _T\} $.
When $j>1 $ this estimation is necessarily conducted using  Monte Carlo  simulations whose results are in turn substituted in the expressions~(\ref{hedge general frequency 1})--(\ref{hedge general frequency 3}) which are also evaluated using Monte Carlo. The variance of the resulting estimator is too high and makes the results unacceptable. Moreover,  the option prices that result from this technique cannot be interpreted as arbitrage free prices if the hedging frequency is lower than periods at which asset prices are observed. Indeed,  in the case $j> 1$, the resulting measure is a martingale measure  for the discounted partial process $\{\widetilde{S}_j, \widetilde{S}_{2j},\ldots, \widetilde{S}_T\}$ but not for the entire discounted price process $\{\widetilde{S}_1, \widetilde{S}_{2},\ldots, \widetilde{S}_T\}$ and therefore there is a leeway for arbitrage.

 As described in Basak and Chabakauri~\citeyearpar{Basak12}, another potential problem with variance minimizing hedging strategies is its time-inconsistency. These authors argue that those can be avoided by either considering that the market is complete or by assuming a zero risk premium for the asset return. The former clearly does not fit our GARCH setting while the latter is equivalent to the  discounted asset price being a martingale under the physical measure; hence, one way to overcome the time-inconsistency problem is to consider hedging directly under a risk neutral measure. 
 
The reasons that we mentioned above lead us to explore the local risk minimization strategy for martingale measures, despite the fact that the historical probability associated with the price process represents a more natural approach  from a risk management perspective. 
We derive below recursive pricing and hedging relations using  risk minimization under risk-neutral measures. These are obtained  in a straightforward manner using expressions~(\ref{hedge general frequency 1})--(\ref{hedge general frequency 3}) combined with the martingale hypothesis in its statement that makes the stochastic discount factors~(\ref{Radon-Nikodym derivative mmm}) $\{N _t\}_{t \in \{j, 2j, \ldots, T  \}} $ automatically all equal to $1$. 

\begin{proposition}\label{HedgingQ}
Let  $Q$ be an  equivalent martingale measure for the price process $\{ S _t\} $ and let $H(S _T)$ be a European contingent claim that depends on the terminal value of the risky asset  $S _t $. The locally risk minimizing strategy every $j$ time steps with respect to the measure $Q$ is determined by the recursions:
\begin{eqnarray}
V _T^{(j)} &= & H,\label{hedge general martingale frequency 1}\\
\xi^{Q}_{t+j} &= & {\rm e}^{-r(T-t)}\frac{E^Q\left[H(S _T)\left(S_{t+j}e^{-rj}-S _t\right) \mid \mathcal{F}_{t}\right]}{{\rm Var}^Q \left[
S_{t+j}{\rm e}^{-rj}-S _t \mid \mathcal{F}_{t}\right]},\label{hedge general martingale frequency 2}\\
V _t ^{(j)}&= &E^Q \left[e^{-r(T-t)}H(S _T) \mid \mathcal{F}_{t}\right],\label{hedge general martingale frequency 3}
\end{eqnarray} 
for all $t=0,j,2j, \ldots, T-j $. 
\end{proposition}
For daily hedging ($j=1 $), the denominator  of~(\ref{hedge general martingale frequency 2}) can be explicitly written down in terms of the $\mathcal{F}_t $-conditional cumulant generating function of $\epsilon_t^\ast  $ under $Q$  and of the risk neutral conditional variance $\sigma^*_{t}$. 
However, for lower hedging frequencies ($j>1 $) this denominator needs to be evaluated by using the  Monte Carlo estimator for the variance.

The recursion formulas require further  explanation. First,  local risk minimization under a martingale measure does not directly determine the arbitrage free price as in the historical probability optimization case. The option price is, a priori, established by assuming that the discounted value process is a martingale under $Q$;~(\ref{hedge general martingale frequency 3}) serves as a verification for this fact. The value process $V^{(j)}_t$ is the same for any $j$, so it does not depend on the hedging frequency.  $V _0 $  represents the necessary  initial investment to setup the local minimizing strategy to replicate $H$. This coincides with the arbitrage free pricing for the payoff $H$ and it depends on the choice of the pricing kernel $Q$. Secondly, the hedging ratios in~(\ref{hedge general martingale frequency 2}) can be computed using standard Monte Carlo techniques by simulating asset price paths under the risk neutral measure considered. In cases where simulation under $Q$ is not very convenient, one can compute both the numerator and the denominator of~(\ref{hedge general martingale frequency 2}) by simulating price paths together with the corresponding  Radon-Nikodym derivative. 
Note that if $Q$ is chosen to be the minimal martingale measure, the discounted value processes from~(\ref{hedge general frequency 3}) and~(\ref{hedge general martingale frequency 3}) coincide, while their corresponding hedging ratios are different. When $Q$ is the pricing measure given by the extended Girsanov principle, the hedging scheme from~(\ref{HedgingQ}) is the same as the one obtained in Elliott and Madan~\citeyearpar{elliott:madan:mcmm} by minimizing the expected discounted risk adjusted squared costs of hedging under the historical measure.
Finally, the optimal hedge ratios from~(\ref{hedge general martingale frequency 2}) minimize the conditional remaining risk and the global risk, since these criteria are equivalent under a martingale measure. Thus, these strategies are also variance optimal. 

\section{Local risk minimization and GARCH diffusion approximations} 
\label{LRMandDelta}

In this section we derive two types of hedging ratios for GARCH options. First, using an idea inspired by a continuous time approach, we  derive local risk minimization hedges based on the assumption that hedging is performed at the same frequency that the underlying is observed and that the time between observations is small. We then compute locally risk minimizing delta hedges for non-Gaussian GARCH diffusion limits and show the relationship with their discrete time counterparts.

\begin{proposition}
\label{delta hedging volatility correction}
Let $H$ be a European contingent product on the underlying price process $\{S _t\}_{t \in {\cal T}}$ generated by a GARCH model. Let  $\Pi^{Q^{min}}_t = \Pi^{Q^{min}}\left(S _t, \sigma_{t+1}^2 \right) = E^{Q^{min}} \left[ e^{r(T-t)}H \mid \mathcal{F}_t \right] $ and $\Pi^{Q}_t = \Pi^{Q}\left(S _t, \sigma_{t+1}^2 \right) = E^{Q} \left[ e^{r(T-t)}H \mid \mathcal{F}_t \right] $be 
associated arbitrage free prices obtained by using the minimal martingale measure $Q^{min}$ and another arbitrary pricing measure $Q$, respectively. If the time between two consecutive observations approaches zero, the following relations hold:
\begin{description}
\item[(i)] The local risk minimization hedge under $P$ can be approximated by:
\begin{equation}
\label{MVP}
\xi^{P}_{t+1} \approx \Delta^{Q^{min}}_{t,S} + VM^{P}_{t+1}  \Delta^{Q^{min}}_{t, \sigma^2} .  
\end{equation}
Here $\Delta^{Q^{min}}_{t,S} = \frac{\partial \Pi^{Q^{min}}_t}{\partial S _t} $ and $\Delta^{Q^{min}}_{t, \sigma^2} = \frac{\partial\Pi^{Q^{min}}_t}{\partial \sigma_{t+1}^2} $ are the sensitivity of the minimum martingale option price with respect to  the asset price and to its conditional variance, respectively. $VM^{P}_{t+1}$ is the vega multiplier under $P$ given by:
\begin{equation}\label{VMP}
VM^{P}_{t+1} = \frac{{\rm Cov}^P\left(S_{t+1} - S_t, \sigma^2_{t+2} - \sigma^2_{t+1} \mid \mathcal{F} _t\right)}{{\rm Var}^P\left[S_{t+1}-S_t\mid \mathcal{F} _t\right]} .
\end{equation}
\item[(ii)] The local risk minimization hedge under $Q$ can be approximated by:
\begin{equation}
\label{MVQ}
\xi^{Q}_{t+1} \approx \Delta^{Q}_{t,S} + VM^{Q}_{t+1}  \Delta^{Q}_{t, \sigma^2} .  
\end{equation}
Here $\Delta^{Q}_{t,S} = \frac{\partial \Pi^{Q}_t}{\partial S _t} $ and $\Delta^{Q}_{t, \sigma^2} = \frac{\partial \Pi^Q_t}{\partial \sigma_{t+1}^2} $ are the sensitivity of the minimal martingale measure option price with respect to  the asset price and to its conditional variance, respectively. $VM^{Q}_{t+1}$ is the vega multiplier under $Q$ given by:
\begin{equation}\label{VMQ}
VM^{Q}_{t+1} = \frac{{\rm Cov}^Q\left(S_{t+1} - S_t, \sigma^2_{t+2} - \sigma^2_{t+1} \mid \mathcal{F} _t\right)}{{\rm Var}^Q\left[S_{t+1}-S_t\mid \mathcal{F} _t\right]} .
\end{equation}
\end{description}
\end{proposition}

The above results can also be obtained using standard minimum variance techniques. The resulting strategies are called minimum variance hedging ratios and have been investigated in both complete and incomplete continuous-time settings (see, for example, Boyle and Emanuel~\citeyearpar{Boyle}, Bakshi \textit{at al.}~\citeyearpar{BCC} and Alexander and Nogueira~\citeyearpar{AN07a} among others).  Indeed, if we consider a delta-neutral portfolio consisting of a long position on a unit of the derivative product and a short one on $\xi_t $ units of the underlying, we can decide on the quantity which should be invested in the underlying asset  by minimizing the one-step ahead portfolio variance. Note that in this case the option price needs to be predetermined based on a pricing measure since this approach provides a criterion  only for the selection of the optimal hedge and not for both the price and the hedge. In our case, (\ref{MVP}) is obtained by minimizing the portfolio variance under $P$ with the option price given by the minimal martingale measure, while (\ref{MVQ}) is found from the same minimization problem under the risk neutral measure $Q$  used to compute the option price. In fact,  ``optimal" solutions similar to (\ref{MVQ}) can be further constructed for any given probability measure, the vega multiplier being the only quantity which changes with respect to this choice. 

The hedging approximation in  (\ref{MVQ}) can be viewed as a correction of the so called static delta hedge proposed by Duan~\citeyearpar{duan:GARCH:pricing} according to which the hedge ratio is the partial derivative of the option price with respect to the underlying. This hedge coincides with those proposed in Proposition~\ref{delta hedging volatility correction} only when either changes in assets and conditional variances are uncorrelated under $Q$, or when $ \Delta^{Q}_{t, \sigma^2}=0$ which is not consistent with the GARCH setting.  Indeed, as Garcia and Renault~\citeyearpar{garcia:renault} pointed out, Duan's~\citeyearpar{duan:GARCH:pricing} result ignores  that the option price in the GARCH context depends on the conditional variance which is itself a function of the asset price. Using this argument, one more approximation of $\xi_t$ can be  obtained if we use:
\begin{eqnarray*}
\sigma^2_{t+2} - \sigma^2_{t+1} \approx \frac{d \sigma^2_{t+1}}{d S_t}\left(S_{t+1} - S_t\right) . 
\end{eqnarray*}
In this case, the vega multiplier under both $P$ or $Q$ becomes, $VM_t =  d \sigma^2_{t+1}/ d S_t $, and therefore   resembles a total derivative formula for the option price with respect to the asset price.\footnote{In an unreported numerical experiment we tested the hedging performance of a total derivative formulation of  (\ref{MVQ}) for an asymmetric GARCH model driven by Gaussian innovations and the results were quite unsatisfactory.} Engle and Rosenberg~\citeyearpar{ER00}  proposed the use of a total derivative formula for hedging discrete time stochastic volatility options. However, instead of computing the Greeks based directly on GARCH implied option prices, their approach was based on a Hull and White~\citeyearpar{HULL1987} Black-Scholes substitution type formula and a vega multiplier based on the average expected volatility.

\subsection{Semi-explicit solutions under the extended Girsanov principle}
\label{HedgingGaussianGARCH}

The goal of this section is to provide a semi-explicit solution for the above local risk minimizing hedging approximation in a general GARCH setting. Since the minimal martingale measure is signed, we focus only on deriving an expression for $\xi^Q_t$ in (\ref{MVQ}) for a special choice of $Q$. More specifically, we shall consider the extended Girsanov principle as our pricing measure candidate. First, we need to describe our models under the physical and risk neutral measures. As we announced above, we now assume a special parametrization for the conditional mean return and the news impact curve. The dynamics under $P$ are given by:
\begin{eqnarray}
y_t   & = &  r + \lambda\sigma_t - \kappa_{\epsilon_t}\left(\sigma_t\right) + \sigma_t \epsilon_t,\;\;\;\;\epsilon_t |\mathcal{F}_{t-1} \sim \mathbf{D}(0,1), \label{retP} \\
\sigma^2_t & = &  \alpha_0 + \alpha_1\sigma^2_{t-1}\left(\epsilon_{t-1} - \gamma \right)^2 + \beta_1\sigma^2_{t-1} .  \label{volP}
\end{eqnarray}
Here $\lambda$ quantifies the market price of risk per unit volatility and $\gamma$ is the leverage effect parameter.  When $\mathbf{D}(0,1) $ is a standard Gaussian distribution, we have that $\kappa^P_{\epsilon_t}(\sigma _t ) = \sigma _t^2/ 2 $, and hence the return dynamics in (\ref{retP}) are identical to the one in Duan~\citeyearpar{duan:GARCH:pricing}.
The basic principle behind the construction of the extended Girsanov principle relies on changing the conditional mean return distribution so that the conditional variance remains unchanged under the new probability measure, denoted here by $Q^{egp}$. This translates into the following risk-neutral representation for the asset returns:
\begin{eqnarray}
y_t & = &  r - \kappa_{\epsilon_t}\left(\sigma_t\right)  + \sigma_t \epsilon^*_t,\;\;\;\;\epsilon^*_t \sim \mathbf{D}(0,1),  \label{Duan Q y}\\
\sigma^2_t & = & \alpha_{0} + \alpha_{1}\sigma^2_{t-1}\left(\epsilon^*_{t-1} - \lambda - \gamma\right)^2 + \beta_{1}\sigma^2_{t-1}. \label{Duan Q sigma}
\end{eqnarray}
Since the conditional law of $\epsilon_t$ under $P$ is the same as that of $\epsilon^*_t$ under $Q^{egp}$, it is straightforward to verify that the discounted asset prices process is a martingale under $Q^{egp}$.
Equations (\ref{Duan Q y})-(\ref{Duan Q sigma}) generalize the local risk neutral valuation relation  to non-Gaussian GARCH models.
We now write down two expressions that we shall need in deriving the Greeks of a European call option that uses the above GARCH model for the underlying asset. Firstly, using~(\ref{Duan Q y})-(\ref{Duan Q sigma}), the time evolution of the prices under $Q^{egp}$ is given by
\begin{equation}
\label{evolution of prices with Q}
S _T=S _te^{r(T-t)-  \sum_{l=t+1}^T  \kappa_{\epsilon_l}\left(\sigma_l\right)+\sum_{l=t+1}^T \sigma_l\epsilon^*_l}.
\end{equation}
Secondly, we let:
\begin{equation*}
P _i:= \alpha _1\left(\epsilon^* _i- \lambda -\gamma \right)^2+ \beta _1, \quad i=0, \ldots, T.
\end{equation*}
A straightforward induction argument  using~(\ref{Duan Q sigma}) implies that:
\begin{equation*}
\sigma_s ^2=\alpha _0A(s,l)+ \sigma _l^2 B(s,l) \quad \mbox{with \quad $l=0, \ldots T-1 $} \quad \mbox{and\quad $s>l $,}
\end{equation*}
where the coefficients are given by:
\begin{equation}
\label{an expression for A and B}
A(s,l):=\prod_{i=l}^{s-1}P _i \left[ \sum_{k=l}^{s-1}\frac{1}{\prod_{i=l}^k P _i}  \right] \quad \mbox{and} \quad B(s,l):=\prod_{i=l}^{s-1} P _i.
\end{equation}

\begin{proposition}
\label{volatility adjusted delta}
For a European call option with strike price $K$ that has~(\ref{Duan Q y})-(\ref{Duan Q sigma}) as the risk-neutralized log-returns generating process, the quantities needed to compute the $Q^{egp}$-locally risk minimization hedging strategy in (\ref{MVQ}) are given by:
\begin{eqnarray}
\Delta^{Q^{egp}}_{t,S} & = & e^{-r(T-t)} E^{Q^{egp}}\left[\frac{S _T}{S _t}\boldsymbol{1}_{\{S _T > K\}}\mid \mathcal{F}_t\right] , 
\label{Delta Hedge} \\ 
\Delta^{Q^{egp}}_{t,\sigma^2} & = & \frac{e^{-r(T-t)}}{2} E^{Q^{egp}}\left[S _T\left(\sum_{l=1}^{T-t}\frac{ B\left(t+l,t+1\right) } {\sigma_{t+l}} \left(\epsilon^*_{t+l} - \kappa^{\prime}_{\epsilon_{t+l}}\left(\sigma_{t+l}\right)\right)\right) \boldsymbol {1}_{\{S_T > K\}} \mid \mathcal{F}_t\right],   \label{Vega Hedge} \\
VM^{Q^{egp}}_{t+1} & = & \frac{\alpha_1\sigma^2_{t+1}}{e^{r}S_t} {\cdot} \frac{ \left(\kappa^\prime_{\epsilon_{t+1}}\left(\sigma_{t+1}\right)\right)^2 - 2(\lambda +\gamma)\kappa^\prime_{\epsilon_{t+1}}\left(\sigma_{t+1}\right) + \kappa^{\prime\prime}_{\epsilon_{t+1}}\left(\sigma_{t+1}\right) - 1}{e^{ \kappa_{\epsilon_{t+1}}\left(2\sigma_{t+1}\right) - 2 \kappa_{\epsilon_{t+1}}\left(\sigma_{t+1}\right) } - 1 }. \label{VMQcf}
\end{eqnarray}
Here $B(\cdot,\cdot)$ is given by (\ref{an expression for A and B}) with  the convention $B(t+1,t+1):=1$.
\end{proposition}
Equation~(\ref{Delta Hedge}) resembles the formula derived in Duan~\citeyearpar{duan:GARCH:pricing} for Gaussian GARCH models.  This relation also holds for any type of asset return process with a general GARCH risk-neutralization of the type given  in~(\ref{retQ})-(\ref{volQ}). Formula~(\ref{Vega Hedge}) provides a general representation for the option vega under a non-Gaussian GARCH model  risk-neutralized via the extended Girsanov principle. When the innovations are Gaussian, this expression reduces to the one given in   Duan~\citeyearpar{Duan00}  for an asymmetric GARCH model.\footnote{The option vega for a GARCH European Call option was given in   Duan~\citeyearpar{Duan00}  without proof or further numerical implementation.} Since there are no closed form results available for the conditional expectations in the thesis of Proposition~\ref{volatility adjusted delta}, we use Monte Carlo simulations for evaluating them. The vega multiplier is expressed in closed-form in (\ref{VMQcf}) and it depends on the first and second derivatives of the conditional cumulant generating function of the driving noise. A much simpler expression for $VM_t$ can be obtained if one wishes to use the total derivative formula for the hedging ratio in (\ref{MVQ}). In this case, the vega multiplier becomes, $VM_{t+1} = d\sigma^2_{t+1}/dS_t = 2\alpha_1\sigma_t\left(\epsilon_t - \gamma\right)/S_t$. We shall not illustrate numerical results based on that expression. Instead, our aim is to investigate the empirical performance of the delta hedge  from (\ref{Delta Hedge}) and the $Q$-local risk minimizing strategy from (\ref{MVQ}) based on the quantities derived in Proposition \ref{volatility adjusted delta}. 


\subsection{Local risk minimization for non-Gaussian GARCH diffusion models}

In this section we investigate the local risk minimization hedging strategies  within the class of stochastic volatility models which are obtained as weak limits of non-Gaussian GARCH models  under both the underlying and risk-neutralized measures. 
Furthermore, we study their relations with the minimum variance hedges derived in Proposition \ref{delta hedging volatility correction}.

The weak convergence of GARCH models to bivariate diffusions where the volatility process and the price process are uncorrelated was first studied by Nelson~\citeyearpar{Nelson} and further extended by Duan~\citeyearpar{MR1457699} for a  more general class of augmented GARCH models, which nests the most commonly used GARCH prescriptions in the literature. These convergence results are typically based on a set of arbitrary assumptions on the model parameters.\footnote{The non-uniqueness of this kind of assumptions leads to the existence of other diffusion limits in the literature (see, for example, Corradi~\citeyearpar{Corradi}, Heston and Nandi~\citeyearpar{heston:nandi} or Kluppelberg \textit{et al.}~\citeyearpar{Claudia}, among others).}  Since the bivariate diffusion limits of GARCH models proposed by Nelson~\citeyearpar{Nelson} and Duan~\citeyearpar{MR1457699} are of widespread use for pricing and hedging derivatives in the continuous time setting, we restrict our attention to the assumptions in those references. In particular, our results are based on the non-Gaussian GARCH diffusion limits derived in Badescu \textit{et al.}~\citeyearpar{BEO13} based on the extended Girsanov principle.

First, we first consider a finer discretization of the model from~(\ref{retP})-(\ref{volP}) by dividing the interval $[0,\dots, T]$ into $nT$ equally spaced points of length $h = 1/n$ and indexed by $k$, $k = 1,2,\cdots,nT$. The log-price process has the following representation on the filtered probability space $\left (\Omega_n, \mathcal{F}_n, \left\{\mathcal{F}_{kh,n} \right \}_{ k = 0,\dots,nT}, P_n \right )$:
\begin{eqnarray}
 \ln S_{kh,n}  - \ln S_{(k-1)h,n}   & = & \left( r + \lambda \sigma_{kh,n} - \kappa_{\epsilon_{kh,n}}\left(\sigma_{kh,n}\right)\right) h+   \sqrt{h} \sigma_{kh,n} \epsilon_ {kh,n},\;\;\;\; \epsilon_{(k+1)h,n}   \sim  \mathbf{D}(0,1) \label{retPd} \\ 
\sigma^2_{(k+1)h,n} - \sigma^2_{kh,n} & = &  \alpha_0(h) + \alpha_1(h) \sigma^2_{kh,n} \left( \epsilon_{kh,n} - \gamma(h) \right)^2 + \left (\beta_1(h) - 1 \right) \sigma^2_{kh,n}.   \label{volPd}
\end{eqnarray}
 Here,  $\epsilon_{kh,n}$ are i.i.d. $\mathbf{D}(0,1)$-distributed random variables. We denote by $M_{j}$ the  raw moments of $\epsilon_{kh,n}$. Note that when when $h=1$, the above equations reduce to  the  NGARCH(1,1) model in~(\ref{retP})-(\ref{volP}).  
Using the following parametric constraints: 
\begin{eqnarray}\label{PC}
\lim_{h \rightarrow 0} \frac{\alpha_1(h)}{h} = \omega_0,\;\;\;\; \lim_{h \rightarrow 0} \frac{1 - \beta_1(h)  -  \alpha_1(h) \left( 1+ \gamma^2(h)\right) }{h} = \omega_1,\;\;\;\;\lim_{h \rightarrow 0} \frac{\alpha^2_1(h)}{h} = \omega_2,\;\;\;\;\lim_{h \rightarrow 0} \gamma(h) = \omega_3 ,
\end{eqnarray}
 Badescu \textit{et al.}~\citeyearpar{BEO13} showed that when $h$ approaches zero  the process from (\ref{retPd})-(\ref{volPd}) converges weakly to the following bivariate diffusion process:
\begin{eqnarray}
d S_t & = & \left( r + \lambda \sigma_t + \frac{1}{2}\sigma^2_t - \kappa_{\epsilon_t}\left(\sigma_t\right)\right)S_t dt + \sigma_t S_t d W_{1t},  \label{ContinuousPy}\\
d \sigma^2_t & = & \left(\omega_0 - \omega_1 \sigma^2_t \right)dt  + \sqrt{\omega_2}(M_3- 2\omega_3) \sigma^2_t d W_{1t} + \sqrt{\omega_2}\sqrt{M_4 - M^2_3 - 1} \sigma^2_t dW_{2t} .  \label{ContinuousPsigma}
\end{eqnarray}
Here, $W_{1t}$ and $W_{2t}$ are two independent Brownian motions on  $\left(\Omega, \mathcal{F}, \left\{\mathcal{F}_t\right\}_{ t \in [0,\dots,T] }, P \right)$. When the GARCH innovations are standard Gaussian distributed, the continuous time approximation from (\ref{ContinuousPy})-(\ref{ContinuousPsigma})  coincides with   the standard GARCH diffusion limit  obtained in Duan~\citeyearpar{MR1457699}. 
Applying the extended Girsanov  transform to the discrete-time model given in~(\ref{retPd})-(\ref{volPd}) leads to the following risk-neutralized dynamics under $Q^{egp}_n$:
\begin{eqnarray}
 \ln S_{kh,n}  - \ln S_{(k-1)h,n}   & = & \left( r - \frac{1}{h}\kappa_{\epsilon_{kh,n}}\left(\sqrt{h}\sigma_{kh,n}\right)\right) h+   \sqrt{h} \sigma_{kh,n} \epsilon^*_ {kh,n},\;\;\;\; \epsilon^*_{(k+1)h,n}   \sim  \mathbf{D}(0,1) \label{retQd}, \\ 
\sigma^2_{(k+1)h,n} - \sigma^2_{kh,n} & = &  \alpha_0(h) + \alpha_1(h) \sigma^2_{kh,n} \left( \epsilon^*_{kh,n} - \sqrt{h}\rho_{kh,n} - \gamma(h) \right)^2 + \left (\beta_1(h) - 1 \right) \sigma^2_{kh,n}.   \label{volQd}
\end{eqnarray}
Here, the driving noise process $\left\{\epsilon^*_{kh,n}\right\}_{k=0,\dots,nT}$ represents a sequence of independent random variables $\mathbf{D}(0,1)$-distributed under $Q^{egp}_n$, which are related to the underlying innovations by $\epsilon^*_{kh,n} = \epsilon_{kh,n} + \sqrt{h} \rho_{kh,n}$, where $\rho_{kh,n}$ is the market price of risk given by:
\begin{eqnarray}
\rho_{kh,n} = \lambda + \frac{  \frac{1}{h} \kappa_{\epsilon_{kh,n}}\left(\sqrt{h}\sigma_{kh,n}\right) - \kappa_{\epsilon_{kh,n}}\left(\sigma_{kh,n}\right) } {\sigma_{kh,n}}. \label{MPR}
\end{eqnarray}
Note that when $h=1$ the price of risk becomes $\lambda$ and the above risk neutralized dynamics reduces to the one in (\ref{Duan Q y})-(\ref{Duan Q sigma}). Badescu \textit{et al.} (2013) showed that under the parametric constraints in (\ref{PC}), the discrete time process  (\ref{retQd})-(\ref{volQd}) weakly converges to the following stochastic volatility model:
\begin{eqnarray}
d S_t & = & r S_t dt + \sigma_t S_t d W^*_{1t},  \label{ContinuousQy}\\
d \sigma^2_t & = & \left(\omega_0 - \left(\omega_1 - 2\sqrt{\omega_2} \omega_3 \rho_t \right) \sigma^2_t \right)dt  + \sqrt{\omega_2}(M_3- 2\omega_3) \sigma^2_t d W^*_{1t} + \sqrt{\omega_2}\sqrt{M_4 - M^2_3 - 1} \sigma^2_t dW^*_{2t} .  \label{ContinuousQsigma}
\end{eqnarray}
Here, $\rho_t$ is the continuous time limit of $\rho_{kh,n}$ in (\ref{MPR}) given by:
\begin{equation*}
 \rho_t = \lambda + \frac{ \frac{1}{2} \sigma^2_t  - \kappa_{\epsilon_t}\left(\sigma_t\right) }{\sigma_t}.
\end{equation*}
The two  Brownian motions $W^*_{1t}$ and $W^*_{2t}$ are independent of  $\left(\Omega, \mathcal{F}, \left\{\mathcal{F}_t\right\}_{ t \in [0,\dots,T] }, Q^{egp} \right)$ and $dW^*_{1t} = dW_{1t} + \rho_t dt$ and  $dW^*_{2t} = dW_{2t} - \rho_t M_3/\sqrt{M_4 - M^2_3 - 1}dt$. When the innovations are standard Gaussian distributed, it follows that $\rho_t = \lambda$ and thus the continuous time limit (\ref{ContinuousQy})-(\ref{ContinuousQsigma})  coincides with the GARCH diffusion limit in Duan~\citeyearpar{MR1457699}. However, for skewed innovations (i.e. $M_3 \neq  0$), the continuous time  implied extended Girsanov principle induces a non-zero market price of $W_{2t}$ risk and the above risk-neutralized dynamics are no longer obtained by applying the the minimal martingale measure to the process (\ref{ContinuousPy})-(\ref{ContinuousPsigma})   (that is, when $dW^*_{2t} = dW_{2t}$). Under this  change of measure, the asset price dynamics is the same as in (\ref{ContinuousQy}), but the risk-neutral conditional variance becomes:
 \begin{equation}
 d \sigma^2_t  =  \left(\omega_0 - \left(\omega_1 + \sqrt{\omega_2} \left( M_3 - 2\omega_3\right)\rho_t \right) \sigma^2_t \right)dt  + \sqrt{\omega_2}(M_3- 2\omega_3) \sigma^2_t d W^*_{1t} + \sqrt{\omega_2}\sqrt{M_4 - M^2_3 - 1} \sigma^2_t dW^*_{2t} .  \label{volMMM}
 \end{equation}
An interesting aspect is that the above limit (\ref{ContinuousQy})-(\ref{ContinuousQsigma}) can be also obtained by applying the exponential affine change stochastic discount factor of Christoffersen \textit{et al.}~\citeyearpar{Chris09b} to the underlying GARCH model. 
\begin{proposition}
\label{lrm SV}
Suppose that the asset returns are governed by the stochastic volatility process determined by (\ref{ContinuousPy})-(\ref{ContinuousPsigma}). Then the locally risk minimizing hedging strategies with respect to $P$ and $Q^{egp}$ are given by:
\begin{eqnarray}
\xi^P_t & = & \Delta^{Q^{min}}_{t,S} + \frac{\sigma_t}{S_t} \sqrt{\omega_2}\left(M_3 - 2\omega_3\right) \Delta^{Q^{min}}_{t,\sigma^2}, \label{xiP} \\
\xi^{Q^{egp}}_t & = & \Delta^{Q^{egp}}_{t,S} + \frac{\sigma_t}{S_t} \sqrt{\omega_2}\left(M_3 - 2\omega_3\right) \Delta^{Q^{egp}}_{t,\sigma^2} . \label{xiQ}
\end{eqnarray}
\end{proposition}
Here the above option Greeks under $Q^{egp}$ and $Q^{min}$ are computed based on  prices associated with asset returns from (\ref{ContinuousQy}) and conditional variance dynamics from (\ref{ContinuousQsigma}) and (\ref{volMMM}), respectively. 
The proof of  Proposition~\ref{lrm SV} follows the same three steps procedure as in Poulsen \textit{et al.}~\citeyearpar{poulsen:2009}; this is based on the algorithm proposed by  El Karoui \textit{et al.}~\citeyearpar{elkaroui:1997} according to which the market is first completed by introducing another tradable asset, then the delta hedge is computed in this setting and finally projected onto the original market.
We notice that the vega multipliers are the same for the two hedging schemes. Moreover,  when the driving noise is  Gaussian distributed, the two local risk minimization problems lead to the same solution since the extended Girsanov principle and the minimal martingale measure have the same GARCH diffusion dynamics associated. More generally, this result also holds true in the presence of arbitrary symmetric GARCH innovations. The use of the partial derivative of the option price with respect to the asset price, $\Delta_{t,S}$,  as the appropriate  hedging ratio within the class of GARCH diffusion models is justified only when either $M_3 = 2\omega_3$ or, in particular, when both quantities are zero. The latter corresponds to the case of a  GARCH underlying model without a leverage effect driven by symmetric innovations.  In general, if we fit an asymmetric GARCH model  to historical data, we obtain a negative leverage effect (i.e. $\omega_3 > 0$) and a negative skewness ($M_3 < 0$). Hence, a non-negative value of the option vega leads to a smaller value of the adjusted hedge ratio in (\ref{xiP}) or (\ref{xiQ}). 

Examining the expressions of the local risk minimizing strategies  based on discrete time GARCH approximations and their diffusion limits, one can establish a relationship between them \footnote{Prigent and Scaillet (2002) investigate the weak convergence of local risk minimization strategies to the standard delta hedges given by the partial derivatives of the option price with respect to the underlying. In particular, they showed such a convergence result when the asset price is modeled via a stochastic volatility diffusion driven by uncorrelated Brownian motions. A key ingredient in their derivation is provided by the Markovian structure of the model, which is only the case in a GARCH(1,1) setup. However, a rigorous proof for such a result is beyond the aim of our paper.}. For example, if standard regularity assumptions hold, which ensure the convergence of the partial derivatives of option prices from GARCH to diffusions, we only need to show such a result for the corresponding vega multipliers. This is illustrated in the result below.
\begin{proposition}
\label{lrm GARCH}
Suppose that $\left(S_{kh,n}, \sigma^2_{(k+1)h,n}\right)$ follows the dynamics of (\ref{retPd})-(\ref{volPd}) under the historical measure $P_{n}$ and (\ref{retQd})-(\ref{volQd}) under the risk neutral measure $Q^{egp}_n$. Then, under the parametric constraints (\ref{PC}), the following limiting results for the vega multipliers defined in (\ref{VMP}) and (\ref{VMQ}) hold:
\begin{equation}
\label{justification}
\lim_{h \to 0} VM^{P_n}_{(k+1)h,n} \left(S,\sigma^2\right)  = \lim_{h \to 0} VM^{Q^{egp}_n}_{(k+1)h,n} \left(S,\sigma^2\right)   = \frac{\sigma}{S} \sqrt{\omega_2}\left(M_3 - 2\omega_3\right).
\end{equation}
Here, vega multipliers are computed conditionally on the filtration  $\mathcal{F}^{\left(S,\sigma^2\right)}_{kh,n} = \mathcal{F}_{kh,n} \cup  \left\{\left(S_{kh,n},\sigma^2_{(k+1)h,n}\right) = \left(S,\sigma^2\right)\right\} $. 
\end{proposition}


\section{Empirical analysis} 
\label{NIGSection}

In this section we assess the empirical performance of the different hedging methods that we theoretically studied above and we compare them with the ones exhibited by more standard approaches such as the Ad-hoc Black-Scholes strategy that is well known for its good performance and ease of implementation (see e.g. Dumas {\it et al.}~\citeyearpar{DumasAdhoc} and references therein). Additionally, we study the impact of the initial conditional variance estimation technique chosen at the time of computing the hedges, as well as the sensitivity of our results with respect to the choice of martingale measure.

\subsection{Dataset description and modeling of the underlying asset}

We use European call options on the S\&P 500 index in order to conduct our study which are well known for their liquidity and high volume of trade. The option price quotes are recorded every Wednesday from Jan 7th, 2004 to Dec 29th, 2004. If a given Wednesday is a holiday, then we consider the next trading day. For a given date, we define the average of the bid-ask quotes as the observed  option price and the closing price as the underlying asset price. The dataset contains a total of 1145 European call options contracts 
which have been obtained after applying similar  selection techniques as those proposed by Dumas {\it et al.}~\citeyearpar{DumasAdhoc}. We  exclude contracts with times to maturity smaller than $7$ and higher than $200$ days as well as moneyness outside the interval $[0.9, 1.1]$. We also set restrictions for the daily volume and daily open interest on our option data in order to eliminate inactive options; more specifically, only options with daily trading volume more than 200 in addition of at least 500 open interest are considered.

We model the log-returns of the S\&P 500 index using the NGARCH(1,1) model under $P$ from (\ref{retP})-(\ref{volP}). All parameters are estimated via maximum likelihood (MLE) using the historical log-returns of the index during the in-sample  period 1992--2003, for which we assume a constant daily interest rate $r = 2.8 \cdot 10^{-5}$. Model estimation is carried out only once and the same model parameters are used in the hedging of all the options in the dataset.  We shall consider two different distributions $\mathbf{D}(0,1) $ for the innovations $\{\epsilon _t \} $ of our model. First, we shall work with the standard Gaussian case, that is, $\mathbf{D}(0,1) = {\bf N}(0,1)$, and then we use a special non-Gaussian GARCH models based Normal Inverse Gaussian (NIG) innovations. The MLE parameters of the Gaussian version of~(\ref{retP})-(\ref{volP}) are:
\begin{equation*}
\alpha_0 = 9.941 \cdot 10^{-7}, \quad \alpha_1 = 0.041, \quad \beta_1 = 0.917, \quad \gamma = 0.863, \quad \mbox{and} \quad \lambda = 0.041.
\end{equation*}
Concerning the NIG innovations, we recall that this distribution has been well-studied in the context of the modeling financial returns by Barndorff-Nielsen~\citeyearpar{Barndorff}. More recently, the performance of GARCH option pricing models based on NIG innovations was investigated by Badescu \textit{et al.}~\citeyearpar{Badescu:option:pricing} for European style options and by Stentoft~\citeyearpar{Stentoft08} for American options, among others. Empirical findings indicate that the NIG density is more appropriate for capturing the negative skewness and  thicker tails generally displayed by asset returns than its Gaussian counterpart; moreover, option prices based on a NIG-GARCH model generally outperform the Gaussian benchmark. We recall that the NIG distribution, $ \mathbf{NIG}(k,a,s,\ell)$, with parameters $k, a, s$ and $\ell$ has  probability density function given by: 
\begin{equation}
f^P(x) = \frac{k}{\pi s}\exp{\left[%
\sqrt{k^2-a^2}+a\left(\frac{x-\ell%
}{s}\right)\right]} \frac{K_1\Big(k\sqrt{1+\big(\frac{%
x-\ell}{s}\big)^2}\Big)}{\sqrt{1+\big(\frac{x-\ell%
}{s}\big)^2}},
\end{equation} 
where $K_1(\cdot)$ represents the modified Bessel function of third kind and index 1. The parameter $k$ measures the kurtosis of the distribution, $a$ the asymmetry, $s$ is the scale parameter, and $\ell$ is the location of the distribution.  The cumulant generating function is given by:
\begin{equation}\label{CGF NIG}
\kappa^P(z) = z\ell + \left( \sqrt{k^2 - a^2} - \sqrt{k^2 - (a+zs)^2} \right) . 
\end{equation}
Thus, the first two conditional historical cumulants are:
\begin{eqnarray*}
E^P[\epsilon_t \mid \mathcal{F}_{t-1} ]  & = & \frac{d \kappa^{P}_{\epsilon_t}(z)}{dz}\Big|_{z=0} = \ell + \frac{as}{\sqrt{k^2 - a^2}} ,\\
{\rm Var}^P[\epsilon_t \mid \mathcal{F}_{t-1} ] & = & \frac{d^2 \kappa^{P}_{\epsilon_t}(z)}{dz^2}\Big|_{z=0} =  \frac{s^2 k^{2}}                         {( \sqrt{k^2 - a^2})^{3}} . 
\end{eqnarray*}
Since the GARCH innovations have mean zero and unit variance we can express the scale and location parameters as functions of $a$ and $k$. This leads to a reduction in the number of NIG parameters which need to be estimated. 

The MLE parameters of the NIG version of the GARCH model from (\ref{retP})-(\ref{volP}) are:
\begin{equation*}
\alpha_0 = 8.665 \cdot 10^{-7}, \quad \alpha_1 = 0.047, \quad \beta_1 = 0.909, \quad \gamma = 0.860, \quad \mbox{and} \quad \lambda = 0.041.
\end{equation*}
The estimated values for the NIG invariant parameters are $k = 1.322$ and $ a =  -0.144$, which produces innovations with skewness -0.250 and kurtosis 4.839.

\subsection{Hedging methods and error estimation}

The performance assessment that we carry out in this section consists of constructing, for each of the options in the dataset, a collection of replicating self-financing portfolios made out of a riskless asset and the underlying. These portfolios are constructed based on the hedges proposed in the paper and are rebalanced weekly or, more specifically, each of the days during the life of the contract in which option prices are available. The performance measure that we shall use for a given hedging strategy is the normalized hedging error that we define as follows: consider an option whose life starts at time $t=0 $ and that expires at  $t = T $. Additionally, suppose that we know its price $V _0 $ at time $t=0 $ and that this contract is hedged at times $\{t _0=0, t _1, \ldots, t _K \}$;  let $\xi:=\{\xi_{t _1}, \xi_{t _2}, \ldots, \xi_{t _{K+1}}\}$ and let $\{S_{t _0}, S_{t _1}, \ldots, S_{t _K}\}$ be the corresponding hedges and prices of the underlying, respectively. The normalized hedging error NHE of this contract is defined as:
\begin{equation*}
{\rm NHE} (\xi):= \frac{\left| H(S _T)- V _0- \sum_{i=0}^K \xi_{t _{i+1}} \cdot \left(S_{t _{i+1}}-S_{t _{i}}\right)\right|}{V _0}
\end{equation*}
that is, the ${\rm NHE} (\xi)$ of a contract with respect of a specific hedging prescription $\xi$  is the absolute value of the difference between the value of the dynamically constructed self financing hedging portfolio and the terminal payoff of the option under consideration, written as a fraction of its price. 

This hedging error measure is a price normalized version of the absolute hedging error used in Bakshi {\it et al.}~\citeyearpar{BCC}. The objective of this modification is to be able to compare hedging errors between options that do not have exactly the same strike and time to maturity. A similar measure was used by Poulsen \textit{et al.}~\citeyearpar{poulsen:2009}. The hedging performances will be reported by computing the average NHEs associated to contracts whose specifications are not identical but whose times to maturity and moneyness belong to a given pair of intervals.

The competing hedging strategies that we put to the test are: 
\begin{itemize}
\item $ \xi = \left(\xi_{t_i}\right)_{i \in \{1,2, \ldots, K+1\} }$, where $\xi_{t}$ is the $Q$-local risk minimization (LRM) hedge computed using equation~(\ref{hedge general martingale frequency 2}); we cal, these strategies ``LRM".
\item $ \Delta_S = \left(\Delta_{t_i,S}\right)_{i \in \{1,2, \ldots, K+1\}}$, where $\Delta_{t_i,S}$ is the instantaneous $Q$-delta hedge proposed by Duan~\citeyearpar{duan:GARCH:pricing} and computed using equation~(\ref{Delta Hedge}); we call these approximation strategies ``Delta". 
\item $ \Delta^{SV} = (\Delta^{SV}_{t_i})_{i \in \{1,2, \ldots, K+1\}}$, where $\Delta^{SV}_{t_i}:=\Delta_{t_i,S}+ VM_t \cdot \Delta_{t_i,\sigma^2}$ is the minimum variance instantaneous delta hedge computed using based on Proposition~\ref{volatility adjusted delta} and obtained as a vega correction of the previous hedge;  we call these approximation strategies ``Delta-SV". 
\end{itemize}
The results obtained with these hedges will be compared with a benchmark provided by the Ad-hoc Black-Scholes method implemented using the option prices in the dataset. We recall that this strategy consists of using the quoted option prices associated to a cross section of contracts for any given date to construct a volatility surface parametrized by strike and time to maturity; this is achieved by regressing the strikes and times to maturity of the available contracts on the  standard Black-Scholes volatilities implied by their prices. As a result of this procedure, each day for which option prices are available we obtain a function that assigns a volatility to each strike and time to maturity; the Black-Scholes delta obtained for a given option out of that volatility  value is the so called Ad-hoc delta which yields the hedging benchmark that we shall be using. We emphasize that the weekly hedging frequency that we shall enforce for all the methods considered in our study is dictated by the fact that the Ad-hoc Black-Scholes benchmark requires option prices in order to be implemented and these are only weekly available.

The Ad-hoc approach motivates two different ways to compute the hedges $ \xi, \Delta_S$, and $\Delta^{SV}$ under consideration. Indeed, the expressions that determine all these quantities involve expectations under the risk neutral measure $Q$ that will be computed using Monte Carlo simulations by producing price paths via the GARCH model specification that we are using. The path simulation is initialized by providing the initial price  $S _0$, as well as initial values of the innovation $\epsilon_0$ and volatility $\sigma _0$. Once the initial volatility $\sigma _0$ has been chosen, the initial innovation $\epsilon_0$ is fully determined by the GARCH prescription using the the log-return $y _0 $ of the underlying, hence the only ingredient that remains to be specified is $\sigma _0$. We explore two possibilities in that respect. First, we consider  Monte Carlo estimation using paths initialized with the GARCH volatility. This strategy consists of initializing the path using the conditional volatility at time $t=0 $ associated to the model under  consideration and historical sample of log-returns of the underlying asset. The implementation of this choice requires knowledge only about the historical prices; additionally, paths are initialized in the same way regardless the specification of the contract that we want to hedge.  Secondly, we propose a Monte Carlo estimation using paths initialized with the Ad-hoc volatility.  This approach relies on initializing the price paths with volatilities adapted to the strike and time to maturity of the option that we want to handle by using the Ad-hoc Black-Scholes volatility obtained out of the regression that we explained above involving the option prices quoted the day in which the hedge is constructed. We emphasize that the second method uses more market information and hence the comparison of Ad-hoc methods with their GARCH volatility based counterparts is not completely fair for that reason. 

As we already mentioned, the hedges $ \xi, \Delta_S $, and $\Delta^{SV}$ under consideration will be computed using a risk neutral measure $Q$. Given that the GARCH model chosen for the underlying makes the associated market incomplete,  the choice of martingale measure $Q$ is an issue and hence it is advisable to study the sensitivity of our results with respect to this point; we do so by considering an exponential affine stochastic discount factor also known as an Esscher transform in addition to the extended Girsanov principle proposed in the previous section. For completeness, we provide below the Radon-Nikodym derivatives of the GARCH models based on $ \mathbf{NIG}(k,a,s,\ell)$ innovations associated to both martingale measures.

The extended Girsanov principle density process is given by:
\begin{equation}\label{RNNIGMCMM}
\frac{dQ^{egp}}{dP} = \exp{\left[{\frac{a\lambda T}{s}}\right]} \prod\limits^T_{t=1} \sqrt{\frac{s^2+ (\epsilon_%
t - \ell)^2}{s^2+ (\epsilon_t + \lambda - \ell)^2}} \frac{K_1\Big(k\sqrt{1+\big(\frac{%
\epsilon_t + \lambda-\ell}{s}\big)^2}\Big)} {K_1\Big(k\sqrt{1+\big(\frac{%
\epsilon_t -\ell}{s}\big)^2}\Big)} .
\end{equation}
For the conditional Esscher transform, denoted by $Q^{ess}$,  the Radon-Nikodym derivative is:
\begin{equation}\label{RNNIGESS}
\frac{dQ^{ess}}{dP} = \exp{\left[ - T \sqrt{k^2 - a^2} \right] }\prod\limits^T_{t=1} \exp {\left[ \hat{\theta}_t \sigma_t
(\epsilon_t - \ell) + \sqrt{k^2 - (a+\hat{\theta}_t \sigma_t s)^2} \right] }. 
\end{equation}
Here, for any $ t \in [0,\cdots,T]$, the Esscher parameter $\hat{\theta}_t$ is the unique solution of the so called martingale equation (i.e. this equation ensures that the discounted asset prices are martingales under $Q^{ess}$) that, in the NIG distribution case, is given by the closed form expression:
\begin{equation}
\hat{\theta}_t = -\frac{1}{2}-\frac{a}{\sigma_t s} -%
\frac{1}{2} \sqrt{\frac{( \kappa^P_{\epsilon_t}(\sigma_t) - (\lambda+\ell)\sigma_t )^2}{\sigma^2_t s^2}    \Big(\frac{4 k^2}{\sigma^2_t s^2+ ( \kappa^P_{\epsilon_t}(\sigma_t) - (\lambda+\ell)\sigma_t )^2  }-1\Big)}.
\end{equation}
Since the hedges that we are interested in require the computation of expectations under a martingale measure $Q$, an important technical point is that the Monte Carlo estimator will be implemented using variance reduction techniques adapted to the martingale character of the discounted prices; indeed, when $Q$ is given by the extended Girsanov principle, we will directly simulate under $Q$ using the corresponding expressions~\eqref{retQ}--\eqref{volQ} and will correct the resulting paths using the empirical martingale simulation (EMS) technique proposed by Duan and Simonato~\citeyearpar{duan:ems1}. When $Q$ is the Esscher measure, it is preferable to simulate under the physical measure and to correct the result using the corresponding Radon-Nikodym derivative~\eqref{RNNIGESS} at the time of computing the expectation.

\subsection{Numerical results}

This subsection presents the numerical results obtained in our empirical study. A first set of computations is contained in Tables~\ref{gaussian errors} and~\ref{NIG errors}. Table~\ref{gaussian errors} (respectively Table~\ref{NIG errors}) contains the average normalized hedging errors associated to  different hedges computed via Gaussian (respectively NIG) GARCH Monte Carlo simulation. Each entry in the table has been computed by averaging the normalized hedging errors committed when handling the options contained in the corresponding moneyness-time to maturity bin. The results in both tables are associated to hedges computed using the extended Girsanov principle martingale measure and Monte Carlo simulations with paths that have been initialized with the GARCH implied instantaneous conditional volatility as well as the Ad-hoc volatility. Additionally, the results associated to the Ad-hoc Black-Scholes benchmark are also reported.

\begin{figure}[htbp]
\centering \subfigure[Mean normalized hedging error for short maturity contracts.  The contracts considered in this figure have times to maturity ranging from 7 to 71 days.]{
     \label{fig:errorsall1}
     \includegraphics[width=14cm,height= 6cm,angle=0]{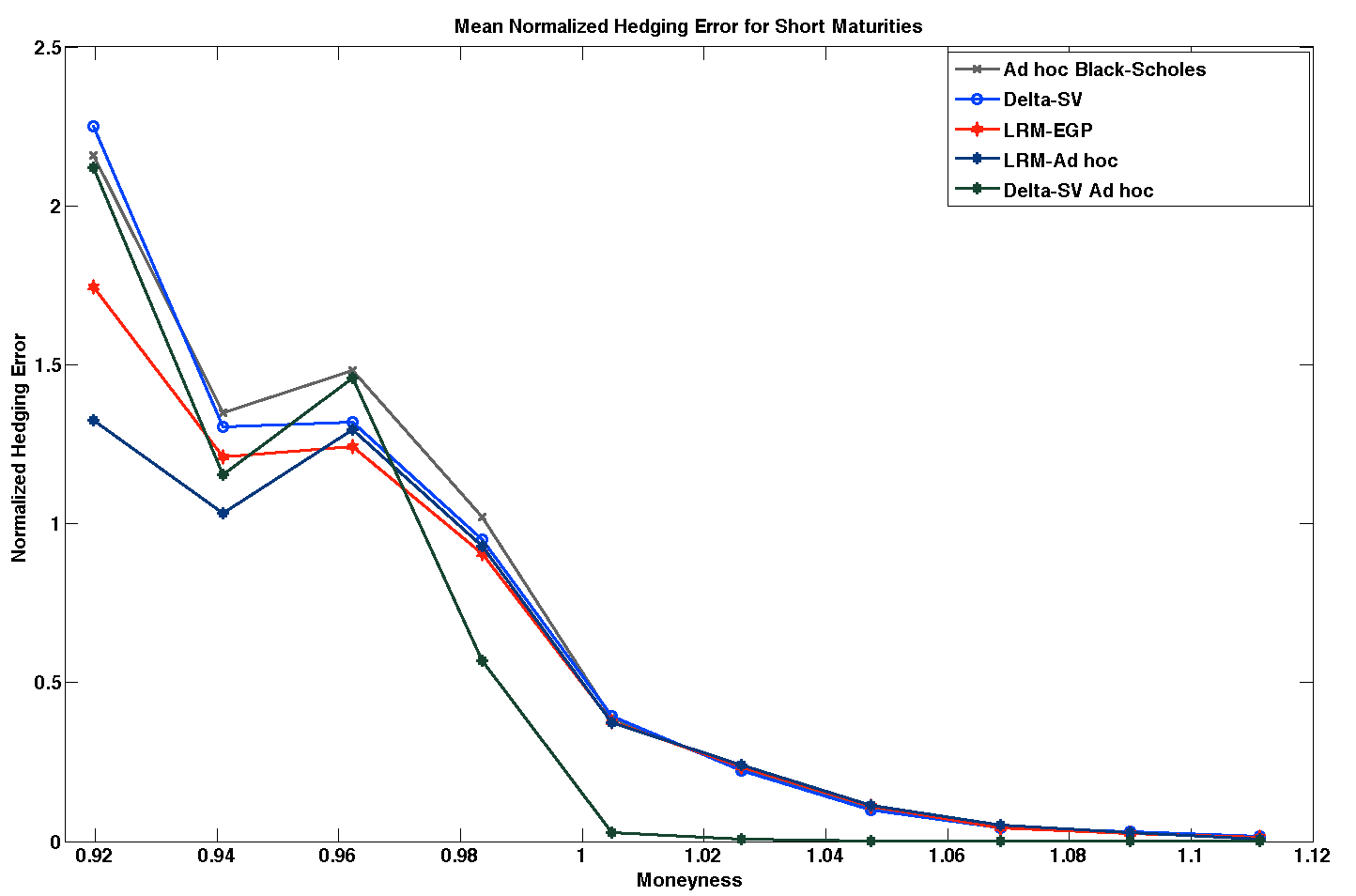}}
  \subfigure[Mean normalized hedging error for medium maturity contracts. The contracts considered in this figure have times to maturity ranging from 71 to 135 days.]{
     \label{fig:errorsall2}
     \includegraphics[width=14.5cm,height=6cm,angle=0]{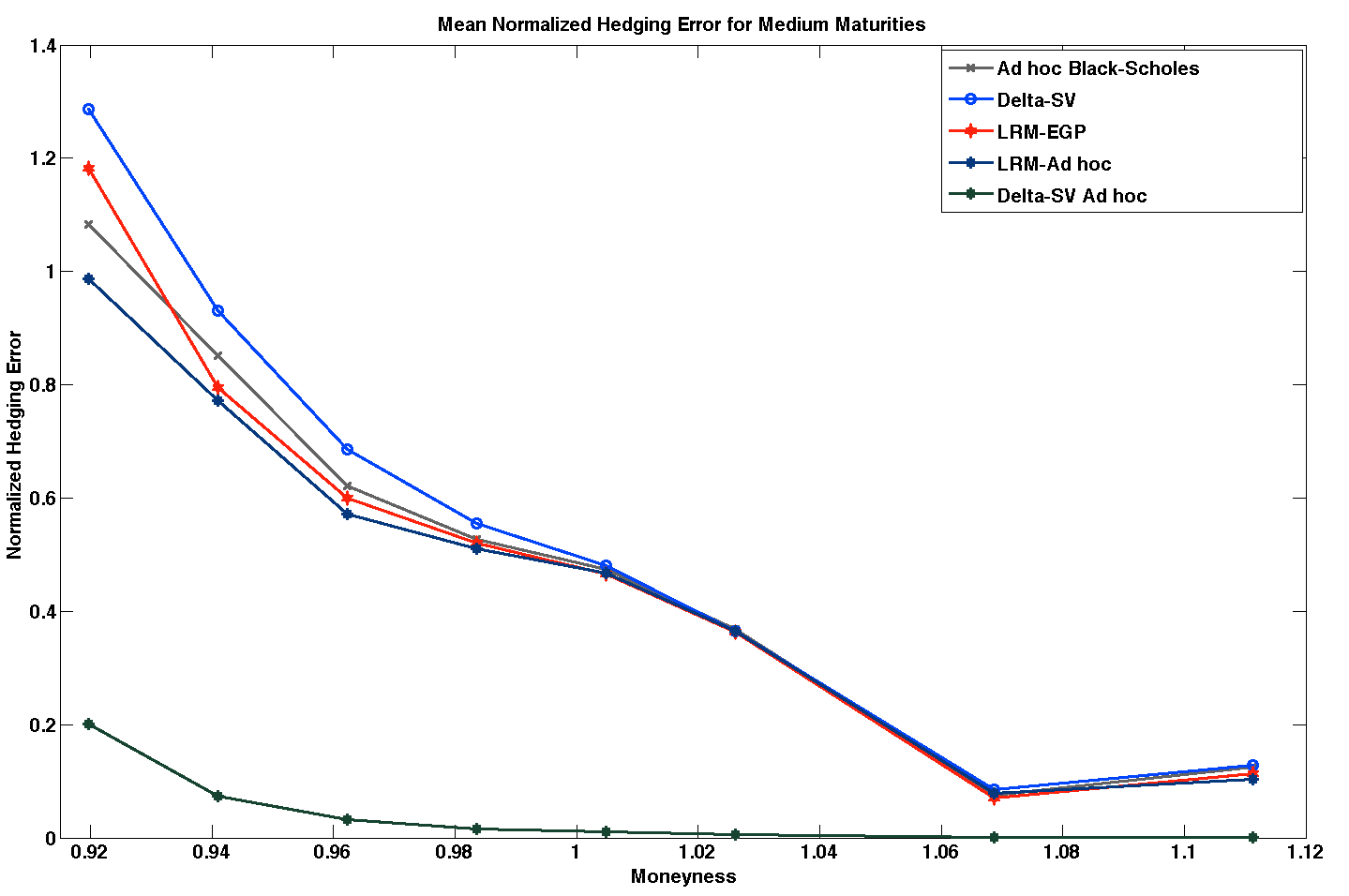}}
  \subfigure[Mean normalized hedging error for long maturity contracts.  The contracts considered in this figure have times to maturity ranging from 135 to 199 days.]{
     \label{fig:errorsall3}
     \includegraphics[width=14.5cm,height=6cm,angle=0]{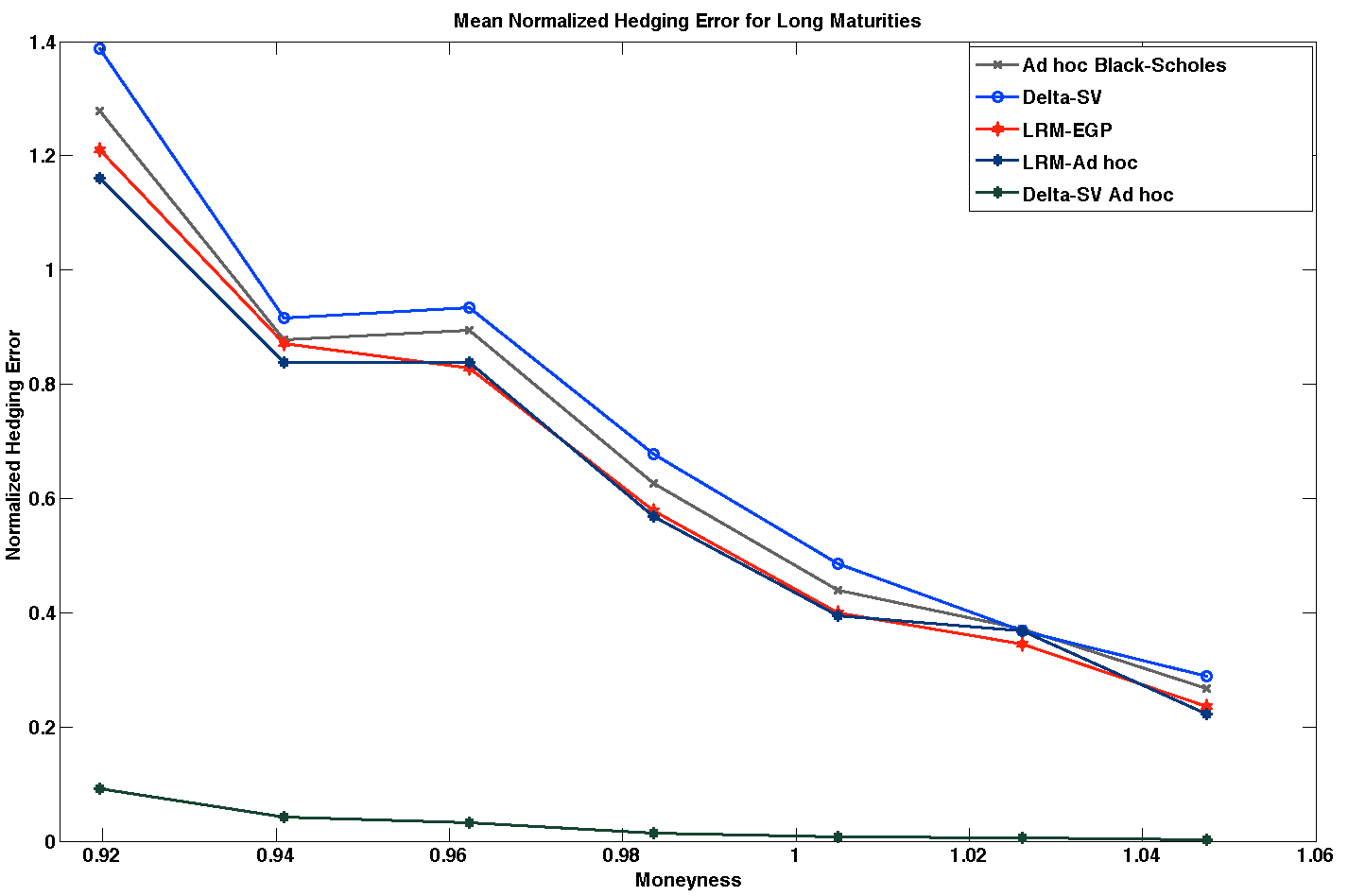}}
\caption{Comparison between the hedging performances obtained using the Ad-hoc Black-Scholes method with the local risk minimizing ratios $\xi $ and the vega corrected hedges $\Delta^{SV} $ computed using NIG-GARCH models for the underlying asset initialized with both the instantaneous GARCH conditional volatility and the Ad-hoc volatility implied by option prices. In contrast with the results presented in the tables~\ref{gaussian errors} and~\ref{NIG errors}, the  average errors reported in these figures have been computed by dividing the moneyness interval $[0.9, 1.1] $ in ten bins (instead of five for the tables).}
\label{Figureall}
\end{figure}

 A first examination of the tables indicates that the local risk minimization methods outperform the Ad-hoc Black-Scholes benchmark for both methods and distributions considered. Regarding delta hedging, this is the case only for the Ad-hoc versions. In general, the hedging schemes computed based on the  Ad-hoc implied volatility starting values consistently outperforms the results obtained using Monte Carlo paths initialized using the GARCH conditional volatility.  In the case of the delta hedges the improvement is only limited for $\Delta_S $ (e.g. the overall average NHE goes from 0.586 to 0.549 for the Gaussian innovations, and from  0.569 to 0.534 for NIG), but outstanding for $\Delta^{SV} $ (e.g. the overall average NHE  goes from 0.572 to 0.204 for Gaussian innovations, and   0.553 to 0.183 for NIG). This is mainly due to the important dependence on initial volatility associated to the vega correction in this hedge. This effect is more pronounced for longer maturity options; for example, in the case of NIG innovations, for maturities between 7 and 71 days the average NHE varies from 0.627 to 0.495, while for  maturities greater than 135 days it varies from 0.674 to 0.026. The same effect can be observed by analyzing the results for each moneyness bin, where the improvement is more pronounced for in-the-money options.  This can be further visualized in the Figure~\ref{Figureall} where we recall that in contrast with the results presented in the Tables~\ref{gaussian errors} and~\ref{NIG errors}, the  average errors reported in the figures have been computed by dividing the moneyness interval $[0.9, 1.1] $ in ten bins,  instead of five for the tables. The vega correction that differentiates the proposed stochastic volatility adjusted delta $\Delta ^{SV} $ and the standard option delta $\Delta_S $ has a positive effect in the empirical hedging performance. For example, in the NIG-GARCH case, the total average NHE goes from 0.569 to 0.553 when the historical GARCH implied variance is used as a starting value.  
 
\begin{figure}[htbp]
\centering \subfigure[Mean normalized hedging error for short maturity contracts.  The contracts considered in this figure have times to maturity ranging from 7 to 71 days.]{
     \label{fig:errorsgaussian1}
     \includegraphics[width=14cm,height= 6cm,angle=0]{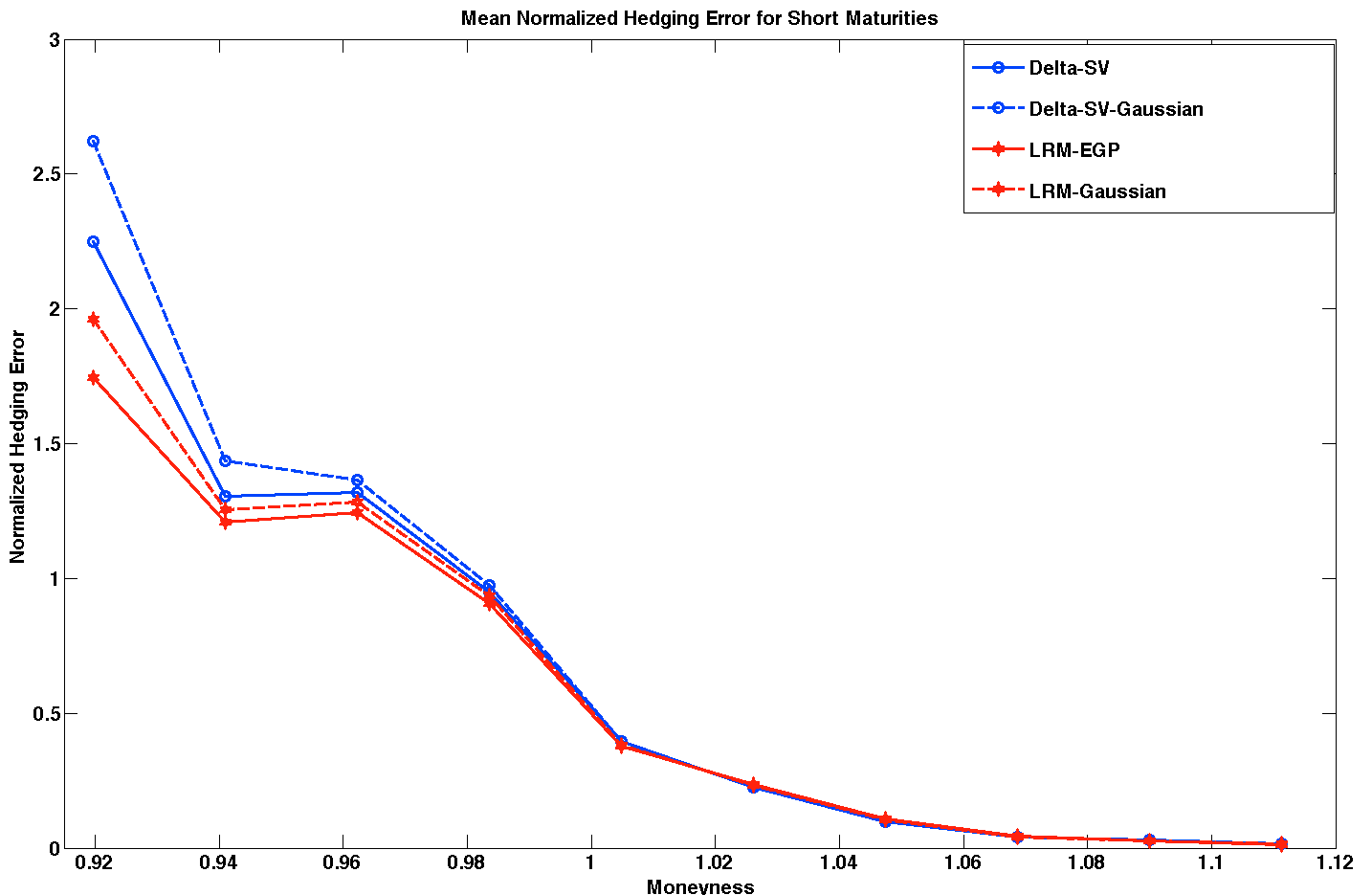}}
  \subfigure[Mean normalized hedging error for medium maturity contracts. The contracts considered in this figure have times to maturity ranging from 71 to 135 days.]{
     \label{fig:errorsgaussian2}
     \includegraphics[width=14.5cm,height=6cm,angle=0]{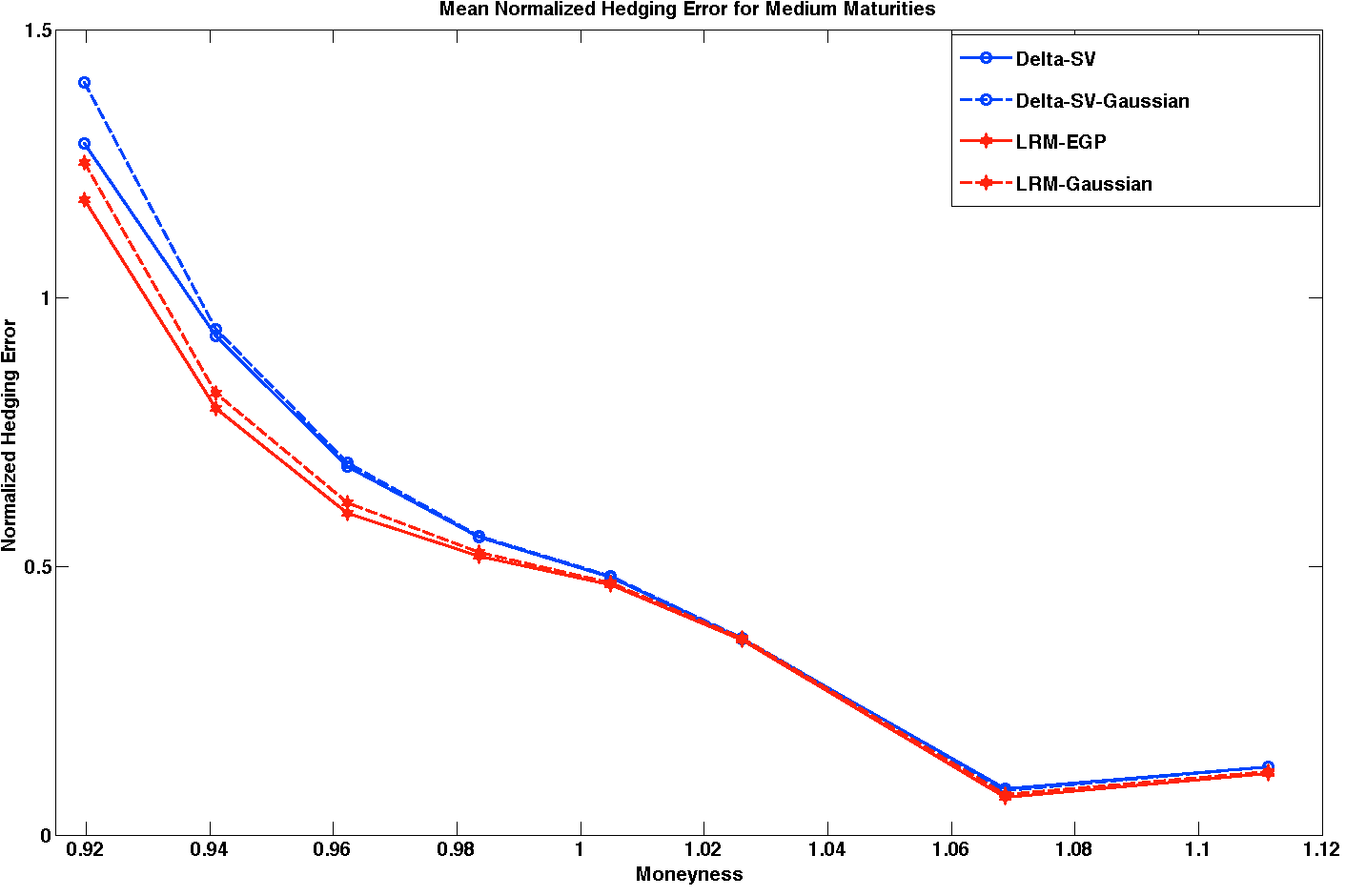}}
  \subfigure[Mean normalized hedging error for long maturity contracts.  The contracts considered in this figure have times to maturity ranging from 135 to 199 days.]{
     \label{fig:errorsgaussian3}
     \includegraphics[width=14.5cm,height=6cm,angle=0]{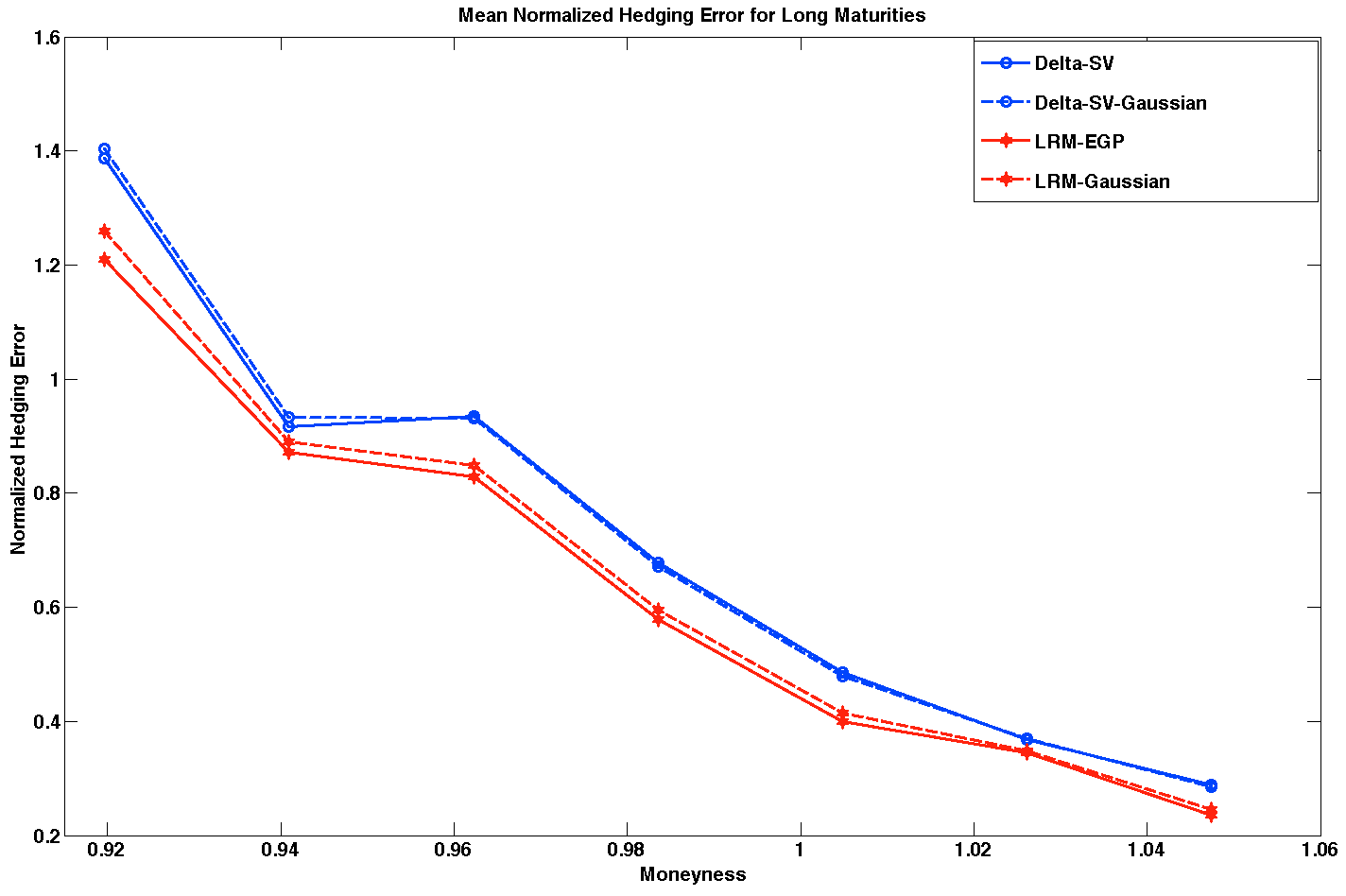}}
\caption{Comparison of the hedging performances obtained using the local risk minimizing ratios $\xi $ and the vega corrected hedges $\Delta^{SV} $ computed using NIG-GARCH and Gaussian GARCH models for the underlying asset. In contrast with the results presented in the tables~\ref{gaussian errors} and~\ref{NIG errors}, the  average errors reported in these figures have been computed by dividing the moneyness interval $[0.9, 1.1] $ in ten bins (instead of five for the tables).}
\label{FigureGaussian}
\end{figure}

Next, we analyze the importance of using different conditional distributions for our GARCH model. The results in Tables~\ref{gaussian errors} and~\ref{NIG errors} indicate that hedges carried out using Gaussian GARCH models exhibit  systematically a lower performance than their NIG counterparts, regardless the method considered and the contract specifications. This fact is further graphically illustrated in the Figure~\ref{FigureGaussian}, where we notice that the differences between the two models are more pronounced for out-of-money options. Thus, the use of a more appropriate skewed and leptokurtic distribution also improves the overall hedging performance. 

 The sensitivity of the hedging performance in the NIG-GARCH case with respect to the choice of pricing measure is very limited. In Table~\ref{Girsanov-Esscher comparison} we have recomputed the local risk minimization $\xi$ and delta $\Delta_S $ hedges using the Esscher transform instead of the extended Girsanov principle that had been used in Tables~\ref{gaussian errors} and~\ref{NIG errors}. As it can be seen, the total average NHE goes from 0.496 for the extended Girsanov principle to 0.492 for the Esscher transform in the case of local risk minimization and from 0.569 to 0.572 for the delta hedges. This analysis is not carried out for the Gaussian case since both martingale measures provides the same risk-neutralized dynamics.

\begin{table}[!htp]
\noindent\makebox[\textwidth]{ 
\begin{tabularx}{1.05\textwidth}{X} 
 \scalebox{0.96}{ 
\begin{tabular}{llcccccc} 
\toprule 
\multicolumn{8}{c}{{\bf NORMALIZED HEDGING ERRORS (Gaussian-GARCH)}} \\
\midrule 
\multicolumn{2}{c}{ }&\multicolumn{5}{c}{{\bf Moneyness $S _0/K $}}&{\bf Average}\\ 
\cmidrule(r){3-7} 
{\bf Method}&{\bf Maturities}&\multicolumn{1}{c}{$[0.91, 0.95]$}&\multicolumn{1}{c}{$[0.95, 0.99]$} 
&\multicolumn{1}{c}{$[0.99, 1.04]$} 
&\multicolumn{1}{c}{$[1.04, 1.08]$} 
&\multicolumn{1}{c}{$[1.08, 1.12]$} 
&{\bf NHE}\\
\midrule 
\multirow{3}{*}{{\bf Ad-hoc BS}} 
&\multirow{1}{*}{${\bf 7}\leq T\leq{\bf 71}$} 
&1.545&1.239&0.364&0.078&0.025&0.650\\
  &\multirow{1}{*}{${\bf 71}\leq T\leq{\bf 135}$} 
&0.944&0.576&0.459&0.075&0.124&0.436\\
  &\multirow{1}{*}{${\bf 135}\leq T\leq{\bf 199}$} 
&1.078&0.742&0.428&0.267&---&0.629\\
  \midrule 
\midrule 
 {\bf Average NHE}&&1.189&0.852&0.417&0.140&0.075&0.535\\
 \midrule 
\midrule 
\multirow{3}{*}{{\bf LRM}} 
&\multirow{1}{*}{${\bf 7}\leq T\leq{\bf 71}$} 
&1.425&1.098&0.361&0.080&0.025&0.598\\
  &\multirow{1}{*}{${\bf 71}\leq T\leq{\bf 135}$} 
&1.005&0.574&0.456&0.076&0.118&0.446\\
  &\multirow{1}{*}{${\bf 135}\leq T\leq{\bf 199}$} 
&1.074&0.705&0.404&0.246&---&0.607\\
  \midrule 
\midrule 
 {\bf Average NHE}&&1.168&0.792&0.407&0.134&0.071&0.515\\
 \midrule 
\midrule 
\multirow{3}{*}{{\bf LRM Ad-hoc}} 
&\multirow{1}{*}{${\bf 7}\leq T\leq{\bf 71}$} 
&1.170&1.134&0.355&0.083&0.023&0.553\\
  &\multirow{1}{*}{${\bf 71}\leq T\leq{\bf 135}$} 
&0.901&0.553&0.455&0.079&0.112&0.420\\
  &\multirow{1}{*}{${\bf 135}\leq T\leq{\bf 199}$} 
&1.034&0.704&0.401&0.242&---&0.595\\
  \midrule 
\midrule 
 {\bf Average NHE}&&1.035&0.797&0.404&0.135&0.067&0.488\\
 \midrule 
\midrule 
\multirow{3}{*}{{\bf Delta}} 
&\multirow{1}{*}{${\bf 7}\leq T\leq{\bf 71}$} 
&1.809&1.210&0.379&0.075&0.029&0.700\\
  &\multirow{1}{*}{${\bf 71}\leq T\leq{\bf 135}$} 
&1.156&0.639&0.471&0.085&0.128&0.496\\
  &\multirow{1}{*}{${\bf 135}\leq T\leq{\bf 199}$} 
&1.182&0.789&0.465&0.289&---&0.681\\
  \midrule 
\midrule 
 {\bf Average NHE}&&1.382&0.879&0.438&0.149&0.078&0.586\\
 \midrule 
\midrule 
\multirow{3}{*}{{\bf Delta Ad-hoc}} 
&\multirow{1}{*}{${\bf 7}\leq T\leq{\bf 71}$} 
&1.494&1.249&0.366&0.076&0.024&0.642\\
  &\multirow{1}{*}{${\bf 71}\leq T\leq{\bf 135}$} 
&1.012&0.605&0.464&0.084&0.122&0.458\\
  &\multirow{1}{*}{${\bf 135}\leq T\leq{\bf 199}$} 
&1.136&0.787&0.458&0.279&---&0.665\\
  \midrule 
\midrule 
 {\bf Average NHE}&&1.214&0.881&0.429&0.146&0.073&0.549\\
 \midrule 
\midrule 
\multirow{3}{*}{{\bf Delta-SV}} 
&\multirow{1}{*}{${\bf 7}\leq T\leq{\bf 71}$} 
&1.740&1.157&0.370&0.076&0.028&0.674\\
  &\multirow{1}{*}{${\bf 71}\leq T\leq{\bf 135}$} 
&1.134&0.627&0.466&0.083&0.128&0.488\\
  &\multirow{1}{*}{${\bf 135}\leq T\leq{\bf 199}$} 
&1.170&0.783&0.460&0.285&---&0.675\\
  \midrule 
\midrule 
 {\bf Average NHE}&&1.348&0.856&0.432&0.148&0.078&0.572\\
 \midrule 
\midrule 
\multirow{3}{*}{{\bf Delta-SV Ad-hoc}} 
&\multirow{1}{*}{${\bf 7}\leq T\leq{\bf 71}$} 
&1.673&1.064&0.025&0.001&0.000&0.553\\
  &\multirow{1}{*}{${\bf 71}\leq T\leq{\bf 135}$} 
&0.150&0.026&0.010&0.001&0.001&0.038\\
  &\multirow{1}{*}{${\bf 135}\leq T\leq{\bf 199}$} 
&0.071&0.023&0.007&0.003&---&0.026\\
  \midrule 
\midrule 
 {\bf Average NHE}&&0.631&0.371&0.014&0.002&0.001&0.204\\
 \midrule 
\bottomrule
\end{tabular}}
\end{tabularx}}
 \caption{Average normalized hedging errors associated to the different hedges computed via Gaussian GARCH Monte Carlo simulation. BS stands for Black-Scholes and LRM for local risk minimization. Delta indicates the hedging scheme associated to the ratio $\Delta_S $ and Delta-SV that of the vega corrected delta $\Delta^{SV}$. When the name of the method contains ``Ad-hoc'', it means that the Monte Carlo have been initialized using the Ad-hoc volatility; otherwise, the initialization is carried out using the instantaneous GARCH conditional volatility. Each entry in the table has been computed by averaging the normalized hedging errors committed when handling the options contained in the corresponding moneyness-time to maturity bin.}
\label{gaussian errors}
\end{table}

\begin{table}[!htp]
\noindent\makebox[\textwidth]{ 
\begin{tabularx}{1.05\textwidth}{X} 
 \scalebox{.96}{ 
\begin{tabular}{llcccccc} 
\toprule 
\multicolumn{8}{c}{{\bf NORMALIZED HEDGING ERRORS (NIG-GARCH)}} \\
\midrule 
\multicolumn{2}{c}{ }&\multicolumn{5}{c}{{\bf Moneyness $S _0/K $}}&{\bf Average}\\ 
\cmidrule(r){3-7} 
{\bf Method}&{\bf Maturities}&\multicolumn{1}{c}{$[0.91, 0.95]$}&\multicolumn{1}{c}{$[0.95, 0.99]$} 
&\multicolumn{1}{c}{$[0.99, 1.04]$} 
&\multicolumn{1}{c}{$[1.04, 1.08]$} 
&\multicolumn{1}{c}{$[1.08, 1.12]$} 
&{\bf NHE}\\
\midrule 
\multirow{3}{*}{{\bf Ad-hoc BS}} 
&\multirow{1}{*}{${\bf 7}\leq T\leq{\bf 71}$} 
&1.545&1.239&0.364&0.078&0.025&0.650\\
  &\multirow{1}{*}{${\bf 71}\leq T\leq{\bf 135}$} 
&0.944&0.576&0.459&0.075&0.124&0.436\\
  &\multirow{1}{*}{${\bf 135}\leq T\leq{\bf 199}$} 
&1.078&0.742&0.428&0.267&---&0.629\\
  \midrule 
\midrule 
 {\bf Average NHE}&&1.189&0.852&0.417&0.140&0.075&0.535\\
 \midrule 
\midrule 
\multirow{3}{*}{{\bf LRM}} 
&\multirow{1}{*}{${\bf 7}\leq T\leq{\bf 71}$} 
&1.334&1.064&0.356&0.083&0.023&0.572\\
  &\multirow{1}{*}{${\bf 71}\leq T\leq{\bf 135}$} 
&0.957&0.561&0.452&0.070&0.114&0.431\\
  &\multirow{1}{*}{${\bf 135}\leq T\leq{\bf 199}$} 
&1.040&0.686&0.390&0.236&---&0.588\\
  \midrule 
\midrule 
 {\bf Average NHE}&&1.110&0.770&0.399&0.130&0.069&0.496\\
 \midrule 
\midrule 
\multirow{3}{*}{{\bf LRM Ad-hoc}} 
&\multirow{1}{*}{${\bf 7}\leq T\leq{\bf 71}$} 
&1.097&1.100&0.354&0.088&0.024&0.533\\
  &\multirow{1}{*}{${\bf 71}\leq T\leq{\bf 135}$} 
&0.858&0.542&0.454&0.078&0.103&0.407\\
  &\multirow{1}{*}{${\bf 135}\leq T\leq{\bf 199}$} 
&0.998&0.686&0.389&0.222&---&0.574\\
  \midrule 
\midrule 
 {\bf Average NHE}&&0.985&0.776&0.399&0.129&0.064&0.471\\
 \midrule 
\midrule 
\multirow{3}{*}{{\bf Delta}} 
&\multirow{1}{*}{${\bf 7}\leq T\leq{\bf 71}$} 
&1.621&1.187&0.382&0.074&0.029&0.659\\
  &\multirow{1}{*}{${\bf 71}\leq T\leq{\bf 135}$} 
&1.105&0.638&0.470&0.088&0.129&0.486\\
  &\multirow{1}{*}{${\bf 135}\leq T\leq{\bf 199}$} 
&1.168&0.795&0.472&0.292&---&0.682\\
  \midrule 
\midrule 
 {\bf Average NHE}&&1.298&0.874&0.442&0.151&0.079&0.569\\
 \midrule 
\midrule 
\multirow{3}{*}{{\bf Delta Ad-hoc}} 
&\multirow{1}{*}{${\bf 7}\leq T\leq{\bf 71}$} 
&1.327&1.228&0.369&0.075&0.024&0.605\\
  &\multirow{1}{*}{${\bf 71}\leq T\leq{\bf 135}$} 
&0.966&0.604&0.464&0.086&0.122&0.448\\
  &\multirow{1}{*}{${\bf 135}\leq T\leq{\bf 199}$} 
&1.122&0.795&0.465&0.284&---&0.666\\
  \midrule 
\midrule 
 {\bf Average NHE}&&1.138&0.876&0.433&0.148&0.073&0.534\\
 \midrule 
\midrule 
\multirow{3}{*}{{\bf Delta-SV}} 
&\multirow{1}{*}{${\bf 7}\leq T\leq{\bf 71}$} 
&1.541&1.122&0.371&0.075&0.028&0.627\\
  &\multirow{1}{*}{${\bf 71}\leq T\leq{\bf 135}$} 
&1.077&0.623&0.465&0.085&0.127&0.475\\
  &\multirow{1}{*}{${\bf 135}\leq T\leq{\bf 199}$} 
&1.154&0.788&0.466&0.289&---&0.674\\
  \midrule 
\midrule 
 {\bf Average NHE}&&1.257&0.844&0.434&0.150&0.078&0.553\\
 \midrule 
\midrule 
\multirow{3}{*}{{\bf Delta-SV Ad-hoc}} 
&\multirow{1}{*}{${\bf 7}\leq T\leq{\bf 71}$} 
&1.401&1.046&0.026&0.001&0.000&0.495\\
  &\multirow{1}{*}{${\bf 71}\leq T\leq{\bf 135}$} 
&0.136&0.026&0.010&0.001&0.001&0.035\\
  &\multirow{1}{*}{${\bf 135}\leq T\leq{\bf 199}$} 
&0.070&0.023&0.007&0.003&---&0.026\\
  \midrule 
\midrule 
 {\bf Average NHE}&&0.536&0.365&0.014&0.002&0.001&0.183\\
 \midrule 
\bottomrule
\end{tabular}}
\end{tabularx}}
 \caption{Average normalized hedging errors associated to the different hedges computed via NIG-GARCH Monte Carlo simulation. BS stands for Black-Scholes and LRM for local risk minimization. Delta indicates the hedging scheme associated to the ratio $\Delta_S $ and Delta-SV that of the vega corrected delta $\Delta^{SV}$. When the name of the method contains ``Ad-hoc'', it means that the Monte Carlo have been initialized using the Ad-hoc volatility; otherwise, the initialization is carried out using the instantaneous GARCH conditional volatility. Each entry in the table has been computed by averaging the normalized hedging errors committed when handling the options contained in the corresponding moneyness-time to maturity bin.}
\label{NIG errors}
\end{table}

\begin{table}[!htp]
\noindent\makebox[\textwidth]{ 
\begin{tabularx}{1.05\textwidth}{X} 
 \scalebox{.96}{ 
\begin{tabular}{llcccccc} 
\toprule 
\multicolumn{8}{c}{{\bf COMPARISON BETWEEN ERRORS COMPUTED USING EGP AND ESSCHER KERNELS}} \\
\midrule 
\multicolumn{2}{c}{ }&\multicolumn{5}{c}{{\bf Moneyness $S _0/K $}}&{\bf Average}\\ 
\cmidrule(r){3-7} 
{\bf Method}&{\bf Maturities}&\multicolumn{1}{c}{$[0.91, 0.95]$}&\multicolumn{1}{c}{$[0.95, 0.99]$} 
&\multicolumn{1}{c}{$[0.99, 1.04]$} 
&\multicolumn{1}{c}{$[1.04, 1.08]$} 
&\multicolumn{1}{c}{$[1.08, 1.12]$} 
&{\bf NHE}\\
\midrule 
\multirow{3}{*}{{\bf LRM-EGP}} 
&\multirow{1}{*}{${\bf 7}\leq T\leq{\bf 71}$} 
&1.334&1.064&0.356&0.083&0.023&0.572\\
  &\multirow{1}{*}{${\bf 71}\leq T\leq{\bf 135}$} 
&0.957&0.561&0.452&0.070&0.114&0.431\\
  &\multirow{1}{*}{${\bf 135}\leq T\leq{\bf 199}$} 
&1.040&0.686&0.390&0.236&---&0.588\\
  \midrule 
\midrule 
 {\bf Average NHE}&&1.110&0.770&0.399&0.130&0.069&0.496\\
 \midrule 
\midrule 
\multirow{3}{*}{{\bf LRM-Esscher}} 
&\multirow{1}{*}{${\bf 7}\leq T\leq{\bf 71}$} 
&1.315&1.052&0.355&0.083&0.023&0.566\\
  &\multirow{1}{*}{${\bf 71}\leq T\leq{\bf 135}$} 
&0.953&0.558&0.452&0.070&0.115&0.430\\
  &\multirow{1}{*}{${\bf 135}\leq T\leq{\bf 199}$} 
&1.036&0.684&0.390&0.230&---&0.585\\
  \midrule 
\midrule 
 {\bf Average NHE}&&1.102&0.765&0.399&0.128&0.069&0.492\\
 \midrule 
\midrule 
\multirow{3}{*}{{\bf Delta-EGP}} 
&\multirow{1}{*}{${\bf 7}\leq T\leq{\bf 71}$} 
&1.621&1.187&0.382&0.074&0.029&0.659\\
  &\multirow{1}{*}{${\bf 71}\leq T\leq{\bf 135}$} 
&1.105&0.638&0.470&0.088&0.129&0.486\\
  &\multirow{1}{*}{${\bf 135}\leq T\leq{\bf 199}$} 
&1.168&0.795&0.472&0.292&---&0.682\\
  \midrule 
\midrule 
 {\bf Average NHE}&&1.298&0.874&0.442&0.151&0.079&0.569\\
 \midrule 
\midrule 
\multirow{3}{*}{{\bf Delta-Esscher}} 
&\multirow{1}{*}{${\bf 7}\leq T\leq{\bf 71}$} 
&1.638&1.195&0.383&0.073&0.028&0.664\\
  &\multirow{1}{*}{${\bf 71}\leq T\leq{\bf 135}$} 
&1.114&0.642&0.472&0.088&0.129&0.489\\
  &\multirow{1}{*}{${\bf 135}\leq T\leq{\bf 199}$} 
&1.176&0.800&0.474&0.294&---&0.686\\
  \midrule 
\midrule 
 {\bf Average NHE}&&1.309&0.879&0.443&0.152&0.079&0.572\\
 \midrule 
\bottomrule
\end{tabular}}
\end{tabularx}}
 \caption{Average normalized hedging errors associated to the local risk minimization $\xi$ and delta $\Delta_S  $ hedges computed via NIG-GARCH Monte Carlo simulation with respect to the pricing measures provided by the extended Girsanov principle and the Esscher transformation. Each entry in the table has been computed by averaging the normalized hedging errors committed when handling the options contained in the corresponding moneyness-time to maturity bin.}
\label{Girsanov-Esscher comparison}
\end{table}

\section{Conclusions}\label{Conc}

In this paper we investigate the hedging of European options when the underlying is modeled with an asymmetric non-Gaussian GARCH model. Since the well-known minimal martingale measure is typically a signed measure for such models, the construction of local risk minimization (LRM) schemes with respect to the historical measure is not possible. Therefore, we first introduce the notion of LRM with respect to a martingale measure for general GARCH setups. When the time between two consecutive hedging trades is small, the local risk minimization can be approximated 
 by minimum variances hedges which depend on the option's delta and vega. Such approximations can be viewed as generalizations of Duan's~\citeyearpar{duan:GARCH:pricing} delta hedge formula by taking into account the dependence of tomorrow's conditional variance on today's asset price. His result is recovered in the case when  asset returns and their conditional variances are uncorrelated.  Other potential hedge ratio candidates which can be obtained using total derivative formulas are discussed. Using the extended  Girsanov principle as our pricing measure, we derive semi-explicit solutions for these minimum variance hedges under a general non-Gaussian framework. Finally, we computed the local risk minimizing hedging strategies for the non-Gaussian GARCH diffusion limits and we showed their relationship with the discrete-time versions. 

An extensive empirical analysis based on S\&P 500 options is provided to test the performance of the proposed hedging schemes. The GARCH models considered are based on Gaussian and Normal Inverse Gaussian innovations. The hedging ratios are computed using Monte Carlo simulation and our schemes are implemented with two different starting conditional variance values: an implied GARCH variance obtained from historical returns and an Ad-hoc Black-Scholes implied variance. Our results indicate that under both scenarios the local risk minimization hedging strategy based on weekly rebalancing outperforms the Ad-hoc Black-Scholes benchmark. Duan's~\citeyearpar{duan:GARCH:pricing}  delta hedge has the worst performance among all methods considered. However, the minimum variance approximation based on the Ad-hoc starting variance is the best hedging scheme with an almost three times smaller hedging error than the Black-Scholes counterpart. Our results are thus consistent with some of the findings from the continuous time hedging literature according to which, local risk minimization provides the best hedge when the two driving Brownian motions of the corresponding stochastic volatility model are correlated. 

Finally, we test the sensitivity of the hedging schemes  with respect to the martingale measure used. Numerical experiments indicate there are no significant differences between the Extended Girsanov Principle and the exponential stochastic discount factor used as the alternative pricing kernel. We found that, for all methods considered, the Normal Inverse Gaussian GARCH model consistently outperforms the Gaussian one.

\newpage

\addcontentsline{toc}{section}{Bibliography}
\bibliographystyle{model2-names}

\newpage

\section{Appendix}
\label{Appendix}

\subsection{Proof of Proposition~\ref{delta hedging volatility correction}}
We prove only the first relationship, the second follows in a similar way. When the time between the observations is small, we can use the following approximation for the change in the option price based on the minimal martingale measure:
\begin{eqnarray}
\label{Option change 1}
\Pi^{Q^{min}}_{t+1} - \Pi^{Q^{min}}_t \approx \frac{\partial \Pi^{Q^{min}}_t}{\partial S _t}\left(S_{t+1} - S_t\right) + \frac{\partial \Pi^{Q^{min}}_t}{\partial \sigma_{t+1}^2}\left(\sigma^2_{t+2} - \sigma^2_{t+1}\right).
\end{eqnarray}
Plugging this into the value process expression from (\ref{hedge general 2}) we have:
\begin{eqnarray*}
\xi^{P}_{t+1} & =  & \frac{{\rm Cov}^P\left(\widetilde{V}_{t+1},\widetilde{S}_{t+1}-\widetilde{S }_t\mid \mathcal{F} _t\right)}                 {{\rm Var}^P\left[\widetilde{S}_{t+1}-\widetilde{S }_t\mid \mathcal{F} _t\right]}  = \frac{{\rm Cov}^P\left(V_{t+1},S_{t+1}\mid \mathcal{F} _t\right)}{{\rm Var}^P\left[S_{t+1}\mid \mathcal{F} _t\right]}\\
& \approx & \frac{{\rm Cov}^P\left(\Pi^{Q^{min}}_t + \frac{\partial \Pi^{Q^{min}}_t}{\partial S _t}\left(S_{t+1} - S_t\right) + \frac{\partial \Pi^{Q^{min}}_t}    {\partial   \sigma_{t+1}^2}\left(\sigma^2_{t+2} - \sigma^2_{t+1}\right), S_{t+1} \mid \mathcal{F} _t\right)}                                          {{\rm Var}^P\left[S_{t+1}\mid \mathcal{F}_t\right]}\\
& = & \Delta^{Q^{min}}_{t,S} + \Delta^{Q^{min}}_{t,\sigma^2}    \frac{{\rm Cov}^P(S_{t+1} - S_t, \sigma^2_{t+2} - \sigma^2_{t+1} \mid \mathcal{F} _t)}     {{\rm Var}^P[S_{t+1}-S _t\mid \mathcal{F} _t]}  =  \Delta^{Q^{min}}_{t,S} + VM^P_{t+1}\Delta^{Q^{min}}_{t,\sigma^2} .\quad \blacksquare
\end{eqnarray*}

\subsection{Proof of Proposition~\ref{volatility adjusted delta}}

We compute separately each term  in the local risk minimization hedging under $Q^{egp}$. Firstly, for any fixed $t$, we assume that certain regularity conditions which allow the interchange between the derivative and the expectation are satisfied. Using~(\ref{evolution of prices with Q}), we can write
\begin{eqnarray*}
\Delta^{Q^{egp}}_{t,S} &=& \frac{\partial \Pi^{Q^{min}}}{\partial S _t} =e^{-r(T-t)}\frac{\partial }{\partial S _t}E^Q_{t}\left[\max\left\{(S _T-K),0 \right\}\right]\notag\\
	&=&e^{-r(T-t)}\frac{\partial }{\partial S _t}E^Q_{t}\left[(S _T-K)\boldsymbol{1}_{\{S _T > K\}}\right]= e^{-r(T-t)}E^Q_{t}\left[\frac{\partial S _T}{\partial S _t}\boldsymbol{1}_{\{S _T > K\}}\right]\notag\\
	&=&e^{-r(T-t)}E^Q_{t}\left[e^{r(T-t)- \frac{1}{2} \sum_{l=t+1}^T \kappa_{\epsilon_l}\left(\sigma_l\right)+\sum_{l=t+1}^T \sigma_l\epsilon^*_l}\boldsymbol{1}_{\{S _T > K\}}\right]=e^{-r(T-t)}E^Q_{t}\left[\frac{S _T}{S _t}\boldsymbol{1}_{\{S _T > K\}}\right].\label{term1 volatility}
\end{eqnarray*}
Second, we compute the option vega $\Delta^{Q^{egp}}_{t,\sigma^2}= \frac{\partial \Pi^{Q^{egp}}}{\partial \sigma_{t+1}^2} $. We start by using~(\ref{an expression for A and B}) in order to rewrite~(\ref{evolution of prices with Q}) as 
\begin{equation*}
S _T=S _t e^{r(T-t)- \kappa_{\epsilon_{t+1}}\left(\sigma_{t+1}\right) + \sigma_{t+1}\epsilon^*_{t+1}- \sum_{l=2}^{T-t} \kappa_{\epsilon_{t+l}}\left(\sqrt{ \alpha _0 A(t+l,t+1)+ \sigma_{t+1}^2B(t+l,t+1)} \right) +\sum_{l=2}^{T-t} \sqrt{ \alpha _0 A(t+l,t+1)+ \sigma_{t+1}^2B(t+l,t+1) }  \epsilon^*_{t+l}}.
\end{equation*}
Using this expression we compute the derivative:
\begin{equation*}
\frac{\partial S _T}{\partial \sigma_{t+1}^2} = \frac{S _T }{2}  \sum_{l=1}^{T-t} \frac{ B(t+l,t+1)}{\sigma_{t+l}} \left(\epsilon^*_{t+l}  - \kappa^\prime_{\epsilon_{t+l}}\left(\sigma_{t+l}\right) \right) ,
\end{equation*}
where we use the convention $B(t+1,t+1):=1 $. We now use the above to compute:
\begin{eqnarray}
\Delta^{Q^{egp}}_{t,\sigma^2}&=& \frac{\partial \Pi^{Q^{egp}}}{\partial \sigma_{t+1}^2} =e^{-r(T-t)}\frac{\partial }{\partial \sigma_{t+1}^2}E^Q_{t}\left[\max\left\{(S _T-K),0 \right\}\right]= e^{-r(T-t)}E^Q_{t}\left[\frac{\partial S _T}{\partial\sigma_{t+1}^2}\boldsymbol{1}_{\{S _T > K\}}\right]\notag\\
	&=&\frac{e^{-r(T-t)}}{2}E^Q_{t}\left[S _T\left( \sum_{l=1}^{T-t} \frac{ B(t+l,t+1)}{\sigma_{t+l}} \left(\epsilon^*_{t+l}  - \kappa^\prime_{\epsilon_{t+l}}\left(\sigma_{t+l}\right) \right)       \right)\boldsymbol{1}_{\{S _T > K\}}\right]. \label{term2 volatility}
\end{eqnarray}
In order to prove the relationship for the vega multiplier in (\ref{VMQcf}), we  rewrite equations (\ref{Duan Q y}) and (\ref{Duan Q sigma}) as:
\begin{eqnarray*}
S_{t+1}  - S_t & = & S_t e^{  r - \kappa_{\epsilon_{t+1}}\left(\sigma_{t+1}\right) + \sigma_{t+1} \epsilon^*_{t+1} }, \\
\sigma^2_{t+2}  - \sigma^2_{t+1} & = & \alpha_0 + \alpha_1\sigma^2_{t+1} \left(\epsilon^*_{t+1} - \gamma - \lambda\right)^2 + \left(\beta_1 - 1\right) \sigma^2_{t+1}.
\end{eqnarray*}
Using the fact that $\kappa_{\epsilon_{t+1}}\left(\cdot\right) = \kappa^*_{\epsilon^*_{t+1}}\left(\cdot\right) $, the denominator becomes:
\begin{eqnarray}
{\rm Var}^{Q^{egp}}\left[S_{t+1} - S_t \mid \mathcal{F}_t\right] & = & S^2_t e^{2\left(r - \kappa_{\epsilon_{t+1}}\left(\sigma_{t+1}\right)  \right)} {\rm Var}^{Q^{egp}}\left[ e^{\sigma_{t+1}\epsilon^*_{t+1}}  \mid \mathcal{F}_t\right] \nonumber \\
& = & S^2_t e^{2r} \left( e^{  \kappa_{\epsilon_{t+1}}\left(2\sigma_{t+1}\right) - 2  \kappa_{\epsilon_{t+1}}\left(\sigma_{t+1}\right) } - 1 \right) . \label{denominator}
\end{eqnarray}
We now compute the numerator:
\begin{eqnarray*}
{\rm Cov}^{Q^{egp}}\left(S_{t+1} - S_t, \sigma^2_{t+2} - \sigma^2_{t+1} \mid \mathcal{F} _t\right)  & = & \alpha_1S_t  e^{r - \kappa_{\epsilon_{t+1}}\left(\sigma_{t+1}\right)  }  \sigma^2_{t+1} {\rm Cov}^{Q^{egp}}\left(e^{\sigma_{t+1}\epsilon^*_{t+1}}, \left(\epsilon^*_{t+1} - \gamma - \lambda\right)^2   \mid \mathcal{F} _t \right) \\
& = & \alpha_1 S_t e^{r - \kappa_{\epsilon_{t+1}}\left(\sigma_{t+1}\right)  }  \sigma^2_{t+1} {\rm Cov}^{Q^{egp}}\left(e^{\sigma_{t+1}\epsilon^*_{t+1}}, \epsilon^{*2}_{t+1} - 2(\gamma+\lambda)\epsilon^*_{t+1}  \mid \mathcal{F} _t \right).
\end{eqnarray*}
Next we express the above in terms the cumulant generating function of the innovations:
\begin{eqnarray*}
{\rm Cov}^{Q^{egp}}\left(e^{\sigma_{t+1}\epsilon^*_{t+1}}, \epsilon^*_{t+1}  \mid \mathcal{F} _t \right) & = & E^{Q^{egp}}\left[ e^{\sigma_{t+1}\epsilon^*_{t+1}} \epsilon^*_{t+1}  \mid \mathcal{F} _t \right] = e^{\kappa_{\epsilon_{t+1}}\left(\sigma_{t=1}\right) } \kappa^\prime_{\epsilon_{t+1}}\left(\sigma_{t+1}\right), \\
{\rm Cov}^{Q^{egp}}\left(e^{\sigma_{t+1}\epsilon^*_{t+1}}, \epsilon^{*2}_{t+1}  \mid \mathcal{F} _t \right) & = & E^{Q^{egp}}\left[ e^{\sigma_{t+1}\epsilon^*_{t+1}} \epsilon^{*2}_{t+1}  \mid \mathcal{F} _t \right] -   E^{Q^{egp}}\left[ e^{\sigma_{t+1}\epsilon^*_{t+1}}  \mid \mathcal{F} _t \right]  \\ & = & e^{\kappa_{\epsilon_{t+1}}\left(\sigma_{t+1}\right) } \left( \left(\kappa^\prime_{\epsilon_{t+1}}\left(\sigma_{t+1}\right) \right)^2 + \kappa^{\prime\prime}_{\epsilon_{t+1}}\left(\sigma_{t+1}\right) - 1 \right).
\end{eqnarray*}
Thus the numerator expression we obtain:
\begin{equation}
{\rm Cov}^{Q^{egp}}\left(S_{t+1} - S_t, \sigma^2_{t+2} - \sigma^2_{t+1} \mid \mathcal{F} _t\right)  = \alpha_1 e^r S_t \sigma^2_{t+1} \left( \left(\kappa^\prime_{\epsilon_{t+1}}\left(\sigma_{t+1}\right) \right)^2 - 2(\gamma+\lambda) \kappa^\prime_{\epsilon_{t+1}}\left(\sigma_{t+1}\right)    +  \kappa^{\prime\prime}_{\epsilon_{t+1}}\left(\sigma_{t+1}\right) - 1      \right) . \label{numerator}
\end{equation}
Substituting (\ref{denominator}) and (\ref{numerator}) into (\ref{VMQcf}), we obtain the desired result for the vega multiplier under $Q^{egp}$. $ \blacksquare$

\subsection{Proof of Proposition~\ref{lrm GARCH}} 

We only derive the limit  for the vega multiplier under $Q^{egp}_n$ for the equation under the physical measure follows in a similar way. For notational convenience we discard the subscripts for the conditional cumulant generating function of  $\epsilon_{(k+1)h,n}$, and we further assume that  all limits are computed conditionally on the enlarged filtration  $\mathcal{F}^{\left(S,\sigma^2\right)}_{kh,n}$. 

 Using  a similar proof as in the previous proposition and  $\alpha^2_1(h) = \omega_2h + o(h)$, the vega multiplier limit reduces to:
 \begin{eqnarray*}
  \lim_{h \to 0} VM^{Q^{egp}_n}_{(k+1)h,n} \left(S,\sigma^2\right)  =  \lim_{h \to 0} \frac{{\rm Cov}^{Q^{egp}_n}\left(S_{(k+1)h,n} - S_{kh,n}, \sigma^2_{(k+2)h,n} - \sigma^2_{(k+1)h,n} \mid \mathcal{F}^{\left(S,\sigma^2\right)}_{kh,n}  \right)}{{\rm Var}^{Q^{egp}_n}\left[S_{(k+1)h,n}-S_{kh,n}  \mid \mathcal{F}^{\left(S,\sigma^2\right)}_{kh,n} \right]} = \sqrt{\omega_2}\frac{\sigma^2}{S}  \lim_{h \to 0} \frac{Q_{(k+1)h,n}}{R_{(k+1)h,n}}. 
 \end{eqnarray*}
Here, $Q_{(k+1)h,n}$ and $R_{(k+1)h,n}$ are functions of $h$ conditionally on $\mathcal{F}^{\left(S,\sigma^2\right)}_{kh,n}$ which  given by: 
\begin{eqnarray*}
Q_{(k+1)h,n}   & = & \sqrt{h}\left( \kappa^\prime\left(\sqrt{h}\sigma_{(k+1)h,n}\right)\right)^2 - 2\left(\sqrt{h}\gamma(h) + h\rho_{(k+1)h,n}\right)\kappa^\prime\left(\sqrt{h}\sigma_{(k+1)h,n}\right) +\sqrt{h}\left( \kappa^{\prime\prime}\left(\sqrt{h}\sigma_{(k+1)h,n}\right) - 1\right) \\
& := & Q_{1,(k+1)h,n}  -  2 Q_{2,(k+1)h,n} +  Q_{3,(k+1)h,n} , \\
R_{(k+1)h,n} & = & e^{ \kappa\left(2\sqrt{h}\sigma_{(k+1)h,n}\right) - 2 \kappa\left(\sqrt{h}\sigma_{(k+1)h,n}\right) } -1.
\end{eqnarray*}
 We first illustrate some useful limits which are needed for computing the above limit ratio. These are based on the second order Taylor expansions around the origin for the cumulant generating functions and its first two derivatives evaluated at  $u = \sqrt{h} \sigma_{(k+1)h,n}$:
\begin{eqnarray*}
\kappa\left( \sqrt{h} \sigma_{(k+1)h,n}\right) = \kappa\left(0\right) +  \sqrt{h} \sigma_{(k+1)h,n} \kappa^\prime\left(0\right) + \frac{h \sigma^2_{(k+1)h,n} }{2}  \kappa^{\prime\prime}\left(0\right) +  o\left(h\right),\\
\kappa^{\prime}\left(\sqrt{h} \sigma_{(k+1)h,n}\ \right) = \kappa^{\prime}\left(0\right) + \sqrt{h} \sigma_{(k+1)h,n}\ \kappa^{\prime\prime}\left(0\right) + \frac{h \sigma^2_{(k+1)h,n} }{2}  \kappa^{(3)}\left(0\right) +  o\left(h\right),\\
\kappa^{\prime\prime}\left(\sqrt{h} \sigma_{(k+1)h,n}\right) = \kappa^{\prime\prime}\left(0\right) + \sqrt{h} \sigma_{(k+1)h,n}\ \kappa^{(3)}\left(0\right) + \frac{h \sigma^2_{(k+1)h,n}}{2}  \kappa^{(4)}\left(0\right) +  o\left(h\right).
\end{eqnarray*}
Using the relationships between the cumulants and the moments of $\epsilon_{(k+1)h,n}$ we derive the following useful limits:
\begin{eqnarray*}
\lim_{h \rightarrow 0} \frac{ \kappa\left(\sqrt{h}\sigma_{(k+1)h,n}\right)} {h} &  = &  \frac{1}{2}\sigma^2,\quad \lim_{h \rightarrow 0} \frac{ \kappa^\prime\left(\sqrt{h} \sigma_{(k+1)h,n} \right)}{\sqrt{h}}  = \sigma, \\
\lim_{h \rightarrow 0} h \rho^\prime_{(k+1)h,n} &  = &   0,\quad \lim_{h \rightarrow 0}  \frac{  \kappa^{\prime\prime}\left(\sqrt{h}\sigma_{(k+1)h,n}\right) - 1}{ \sqrt{h} } = M_3\sigma. 
\end{eqnarray*}
Here, $ \rho^\prime_{(k+1)h,n}$ denotes the derivative of the market price of risk with respect to $h$. We now apply l'H\^{o}pital's rule for calculating the limit. After some tedious algebraic manipulations, the limits of the derivatives of the three quantities in $Q_{(k+1)h,n}$ are given by:
\begin{equation}
\lim_{h \rightarrow 0} Q^\prime_{1,(k+1)h,n}  =  0, ,\quad
\lim_{h \rightarrow 0} Q^\prime_{2,(k+1)h,n}  =  \omega_3\sigma, \quad
\lim_{h \rightarrow 0} Q^\prime_{3,(k+1)h,n}  =  M_3\sigma. \label{L1}
\end{equation}
Similarly, we obtain $\lim_{h \rightarrow 0} R^\prime_{(k+1)h,n}  = \sigma^2$. Substituting this result and (\ref{L1}) into the vega multiplier limit we have:
\begin{equation*}
\label{justification}
 \lim_{h \rightarrow 0} VM^{Q^{egp}_n}_{(k+1)h,n} \left(S,\sigma^2\right)  = \sqrt{\omega_2}\frac{\sigma^2}{S}  \lim_{h \to 0} \frac{Q_{(k+1)h,n}}{R_{(k+1)h,n}}  \frac{\sigma}{S} \sqrt{\omega_2}\left(M_3 - 2\omega_3\right) = \sqrt{\omega_2}\frac{\sigma^2}{S}  \lim_{h \to 0} \frac{Q^\prime_{(k+1)h,n}}{R^\prime_{(k+1)h,n}} =  \frac{\sigma}{S} \sqrt{\omega_2}\left(M_3 - 2\omega_3\right). \blacksquare
\end{equation*}

\end{document}